\documentclass[
superscriptaddress,
preprint,
prd,tightenlines,showpacs,nofootinbib,
eqsecnum,
amsfonts,amsmath,amssymb]{revtex4-1}

\usepackage{bm}
\usepackage{graphicx} 
\usepackage{hyperref}
\usepackage{mathrsfs}
\usepackage{multirow}

\newcommand{\bfy}{\mathbf{y}}
\newcommand{\bfx}{\mathbf{x}}

\newcommand{\bfS}{\mathbf{S}}
\newcommand{\bfSigma}{\mathbf{\Sigma}}
\newcommand{\ov}[1]{\overline{#1}}

\newcommand{\ud}{\mathrm{d}}
\newcommand{\uD}{\mathrm{D}}

\newcommand{\calO}{\mathcal{O}}
\newcommand{\calF}{\mathcal{F}}
\newcommand{\calL}{\mathcal{L}}
\newcommand{\ph}[1]{\phantom{#1}}
\newcommand{\vph}[1]{\vphantom{#1}}

\newcommand{\be}{\begin{equation}}
\newcommand{\ee}{\end{equation}}

\newcommand{\exch}{1\leftrightarrow 2}
\newcommand{\nn}{\nonumber}

\usepackage{ulem}
\usepackage[usenames]{color}

\allowdisplaybreaks

\begin{document}

\title{Cubic order spin effects in the dynamics and gravitational wave energy flux of compact object binaries}

\author{Sylvain \textsc{Marsat}}\email{smarsat@umd.edu}
\affiliation{Maryland Center for Fundamental Physics \& Joint
  Space-Science Center,\\ Department of Physics, University of
  Maryland, College Park, MD 20742, USA}
\affiliation{Gravitational Astrophysics Laboratory, NASA Goddard Space Flight Center, Greenbelt, MD 20771}

\date{\today}

\begin{abstract}
We investigate cubic-in-spin effects for inspiralling compact objects binaries, both in the dynamics and the energy flux emitted in gravitational waves, at the leading post-Newtonian order. We use a Lagrangian formalism to implement finite-size effects, and extend it at cubic order in the spins, which corresponds to the octupolar order in a multipolar decomposition. This formalism allows us to derive the equation of motion, equations of precession for the spin, and stress-energy tensor of each body in covariant form, and admits a formal generalization to any multipolar order. For spin-induced multipoles, i.e. in the case where the rotation of the compact object is sole responsible for the additional multipole moments, we find a unique structure for the octupolar moment representing cubic-in-spin effects. We apply these results to compute the associated effects in the dynamics of compact binary systems, and deduce the corresponding terms in the energy loss rate due to gravitational waves. These effects enter at the third-and-a-half post-Newtonian order, and can be important for binaries involving rapidly spinning black holes. We provide simplified results for spin-aligned, circular orbits, and discuss the quantitative importance of the new contributions.
\end{abstract}


\maketitle


\section{Introduction}\label{Introduction}

The advent of a new generation of ground-based gravitational wave detectors such as advanced LIGO~\cite{ligo} and advanced Virgo~\cite{virgo}, complemented by GEO~\cite{geo} and in a few years KAGRA~\cite{kagra}, will bring gravitational wave astronomy in its observational era. Further on the way, space-based detectors such as eLISA~\cite{elisa13} will enlarge the scope of the observations by targeting a lower frequency band. The most promising sources for both classes of detectors are the inspirals and coalescence of compact object binaries, neutron stars and/or stellar mass black holes for ground-based detectors and massive black holes for the space-based ones.

Data analysis techniques require an accurate modelling of the expected gravitational wave signals, to ensure detection in the search pipelines and to reduce systematic biases on parameter estimation in subsequent analyses (see e.g.~\cite{Cutler+93,CF94}). This has driven a lot of effort to improve the theoretical prediction for the radiation emitted by a binary system in general relativity. Among the different and complementary approaches to the problem, the post-Newtonian theory (abbreviated throughout the paper as PN) allow to cover the long inspiral phase with an analytic expansion~\cite{Bliving}, and also serve as a basis for the effective-one-body inspiral-merger-ringdown templates~\cite{BD99}.

Astrophysical observations suggest that black holes can be commonly close to extremally spinning with $\chi \lesssim 1$ (see e.g.~\cite{Gou+11,Nowak+12} for stellar mass black holes and~\cite{Brenneman+11,Reynolds13} for supermassive black holes), while neutron stars are expected to have limited spins, with $\chi \leq 0.1$ (of the order of $0.4$ for the fastest known millisecond pulsar~\cite{Hessels+06}). Including the effects of the spins\footnote{We use ``spin'' to denote the angular momentum of the compact objects, in a purely classical sense.} in templates is therefore important, at least for black holes, and higher orders in spin beyond the linear level can be significant.

The purpose of this paper is to continue a serie of previous works~\cite{FBB06,BBF06,BFH12, MBFB13, BMFB13, BMB13, MBBB13}, extending the knowledge of spin effects in the dynamics and gravitational waves emitted by such compact object binaries, by computing the cubic-in-spin terms that enter the problem at the 3.5PN order. To this end, we extend to the octupolar order a previously proposed Lagrangian formalism to represent spin-induced finite size effects for compact objects in general relativity.

The study of the motion of compact objects with spin has been ongoing since the seminal works of Mathisson~\cite{Mathisson37, Mathisson40,Mathisson37repub} and Papapetrou~\cite{Papapetrou51spin}, that established the form of the equations of motion and precession at linear order in the spins. Two main routes have been explored to generalize their results and to cover finite-size effects, including the spin-induced ones, that is when the compact object is deformed by its rotation, which enter at quadratic and higher order in spin. In the first route, the object is treated as a point particle with a stress-energy tensor that takes the form a gravitational ``skeleton'', with coefficients that, after some redefinitions, are recast in the form of a spin tensor and higher multipoles. This approach has been developed notably by Tulczyjew and Dixon (see e.g.~\cite{Tulczyjew59,Dixon64,Dixon73,Dixon74,Dixon79,SP10} for a modern application of this approach at quadrupolar order). The second route, pioneered by Hanson and Regge~\cite{HR74} for special relativity (see also the earlier works~\cite{GW67,W69}), consists in introducing angular degrees of freedom for the point particle in a Lagrangian formalism. A quite general treatment of spinning particles in general relativity, including the generalized action described in section~\ref{sec:formalism}, has been proposed by Bailey and Israel~\cite{BI75}. This Lagrangian approach has been subsequently developed in the framework of the effective field theory (hereafter EFT) approach to binary systems in~\cite{Porto06}. It has also been used to build the Hamiltonian of a spinning test particle in a background spacetime~\cite{BRB09}, and in the context of the reduced Hamiltonian approach to the two-body problem in the Arnowitt-Deser-Misner (ADM) formalism~\cite{Steinhoff11}.

Once the formalism for spinning extended bodies is established, it can be specialized to the case where the additional structure of the particle is induced by its spin only. Other finite-size effects such as tidal effects can also be represented with an effective action~\cite{GR06,BDF12}, but for compact objects the effacement principle ensures that they will be formally of the 5PN order~\cite{Damour83houches} and, although important in the last stages of binary neutron stars inspirals, we leave them aside in this paper. The spin-induced quadrupolar moment, proportional to the square of the spin, has been introduced in the context of binary systems in~\cite{Poisson97}, and covariantly generalized for the EFT~\cite{PR08b} and ADM~\cite{Steinhoff11} approaches, as well as in the harmonic coordinates framework that we are using~\cite{BFH12}. At cubic order in spin, only recently has the leading-order coupling been written down in the EFT framework in~\cite{LS14b}, which contains also a similar study at the quartic-in-spin level. The spin-induced moments are proportional to constants, one at each multipolar order, that describe in general the internal structure of the compact object. They can be determined analytically for Kerr black holes and the quadrupolar one has been computed numerically for neutron star models~\cite{LP97}.

The results of this formalism can then be used in the context of the post-Newtonian approximation for binary systems. The derivation of the spin effects both in the dynamics and gravitational waves emitted has a long history, and we give here only an overview of recent results\footnote{Other modern PN results not detailed here are due to the DIRE formalism~\cite{W05} and to the surface-integral approach~\cite{IF03}.}, using (here and througout the paper) the following abbreviations: NS stands for the no-spin terms, SO for spin-orbit (or linear-in-spin), SS for quadratic-in-spin and SSS for cubic-in-spin. In the ADM approach, the NS dynamics has been computed\footnote{With one remaining constant being determined by an analytic self-force calculation~\cite{BD13a}.} to the 4PN order~\cite{JS12,JS13,DJS14}, the SO dynamics to the 3.5PN order~\cite{DJSspin,HS11so,HSS13}, and the SS dynamics to the 3PN order~\cite{HSS10} for the self terms $S_{1}^{2}$, $S_{2}^{2}$ and to the 4PN order~\cite{HS11ss} for the cross terms $S_{1}S_{2}$. In the EFT approach, the NS dynamics has been recovered at 3PN in~\cite{FS11}, the formalism for the inclusion of spins has been tackled in~\cite{Porto06}, and the SO 2.5PN order~\cite{Porto10,Levi10}, the $S_{1}^{2}$, $S_{2}^{2}$ 3PN order~\cite{PR08b} and the $S_{1}S_{2}$ 3PN~\cite{PR08a,Levi08} and 4PN order~\cite{Levi12} have been obtained. The non-trivial consistency with the ADM results has been investigated for the quadratic order in spin in~\cite{HSS12,LS14a,Levi08,Levi10}. The EFT approach can also be applied to the gravitational wave generation, and partial results have been given for the spin terms in the 3PN energy flux~\cite{PRR10} and the 2.5PN waveform~\cite{PRR12}. Finally, approach in harmonic coordinates approach that we follow in this paper and the associated multipolar post-Newtonian wave generation formalism (see~\cite{Bliving} for a review), the dynamics has been obtained to the NS 3.5PN order~\cite{BF00eom,BF01eom,ABF01,BDE04,NB05}, the SO 3.5PN order~\cite{FBB06,MBFB13,BMFB13}, and the SS 2PN order~\cite{BFH12}, together with the energy flux at the NS 3.5PN order~\cite{BIJ02,BFIJ02,BDEI05}, the SO 4PN order~\cite{BBF06,BMB13,MBBB13}, and the SS 2PN order~\cite{BFH12}, and finally the waveform amplitude at the NS 3PN order~\cite{BFIS08} (and at 3.5PN order for the $22$ mode~\cite{FMBI12}), the SO 1.5PN order~\cite{BBF06,ABFO09}, and the SS 2PN order~\cite{BFH12}. The completion of the SS dynamics and energy flux to the 3PN order has been recently achieved~\cite{BFMP15}. Another class of results supplementing the classical PN ones are the effects of tidal heating and torquing of black holes, or absorbed fluxes through the horizon (see e.g~\cite{TMT97,TP08,CPY12}). These effects enter at the 4PN order in the non-spinning case, but at 2.5PN when spins are present, and yield a contribution to the phase evolution of the binary.

Beyond the quadratic order in spin, the leading order cubic and quartic in spin Hamiltonians were obtained for Kerr black holes in~\cite{HS07,HS08}, using an indirect method\footnote{Namely, by constraining the terms allowed in the Hamiltonian by imposing the test-mass limit of a particle around orbiting a Kerr black hole, and by requiring the completion of the Poincar\'e algebra}. The recent work~\cite{LS14b} used the EFT approach to obtain a generalization of these for a general compact object (and corrected the results at quartic order). Our work confirms the SSS part of the results of~\cite{LS14b} for the dynamics, and complements them with the computation of the gravitational wave energy flux. Another recent article recovered partially these results for the dynamics~\cite{Vaidya14}, using a different flavor of the EFT approach. 

The paper is organized as follows. We first lay down the general formalism for extended particles with spin at the octupolar order, introducing couplings to the Riemann tensor and its derivative, in section~\ref{sec:formalism}. In section~\ref{sec:SSC}, we work out the consequences of enforcing the spin supplementary condition, while in section~\ref{sec:spininduced}, we investigate the structure of the quadrupolar and octupolar moments in the spin-induced case. Section~\ref{sec:PNdynamics} presents the necessary formalism for PN calculations and gives the results for the dynamics, and finally section~\ref{sec:PNflux} does the same for the emitted energy flux and contains a discussion of the quantitative importance of the new contributions in the case of circular spin-aligned orbits. We present a generalization of the formalism to higher multipolar order in appendix~\ref{app:generalization}, and compare our results with the ones obtained for the dynamics in the ADM and EFT methods in appendix~\ref{app:ADMEFT}.

Throughout the paper, 1PN order corresponds to $v^{2}/c^{2}\sim G m / r c^{2}$, and we use the notation $\calO(n) = \calO(1/c^{n})$. We adopt the $(-+++)$ sign convention for the metric, and we scale the spin variables as $S = c S_{\rm true} = G m^{2} \chi$, with $\chi$ the dimensionless spin parameter comprised between 0 and 1. Greek indices stand for spacetime indices, and latin indices will represent spatial, Euclidean indices. We will set $G=c=1$ in the first part of the paper, and restore them in the sections~\ref{sec:PNdynamics} and ~\ref{sec:PNflux} where PN results are presented.


\section{Lagrangian formalism for spinning particles}\label{sec:formalism}

In this section, we present the general Lagrangian formalism for spinning point particles, including finite size effects up to the octupolar order. The early and quite general treatment by Bailey and Israel~\cite{BI75} of spinning particles in general relativity already contains many of the results presented here, but we choose to give a comprehensive presentation.


\subsection{Definitions}

We shall first lay down the necessary definitions for our Lagrangian formalism for point particles with spins. The particle worldline will be parametrized by $\tau$, and described by the coordinates $z^{\rho}(\tau)$ and the 4-velocity $u^{\mu} \equiv \ud z^{\mu}/\ud \tau$. At the level of the Lagrangian, $\tau$ will be a freely specifiable parameter and $u^{\mu}$ will not be normalized; at the level of the equations of motion, after variation of the action, we will choose for $\tau$ the proper time parameter with $u_{\mu}u^{\mu} = -1$. We keep the same notation $\tau$ in both contexts.

The central idea of this Lagrangian formalism for particles with spin, introduced by Hanson and Regge in the framework of special relativity~\cite{HR74}, is to represent the rotational degrees of freedom of the compact object using an orthonormal tetrad $\epsilon_{A}^{\ph{A}\mu}(\tau)$ attached to the worldline, and to relate it to a background tetrad field $e_{a}^{\ph{a}\mu}(x)$. Both are orthonormal and verify (with $\bm{\eta}$ the Minkowski metric)
\begin{align}
	e_{a}^{\ph{a}\mu}e_{b\mu} = \eta_{ab} \,, \quad e_{a}^{\ph{a}\mu}e^{a\nu} = g^{\mu\nu} \,, \nn \\
	\epsilon_{A}^{\ph{A}\mu}\epsilon_{B\mu} = \eta_{AB} \,, \quad \epsilon_{A}^{\ph{A}\mu}\epsilon^{A\nu} = g^{\mu\nu} \,.
\end{align}
The tetrad indices $a,A = 0,1,2,3$ are raised and lowered with the Minkowski metric $\bm{\eta}$. The two tetrads are related by Lorentz matrices $\Lambda_{A}^{\ph{A}a}(\tau)$ according to the relations
\begin{align}\label{eq:defLambda}
	e_{a}^{\ph{a}\mu}(z(\tau)) &= \Lambda^{A}_{\ph{A}a}(\tau) \epsilon_{A}^{\ph{A}\mu}(\tau) \,, \\
	\Lambda_{A}^{\ph{A}a} \Lambda_{Ba} &= \eta_{AB} \,, \quad \Lambda_{Aa} \Lambda^{A}_{\ph{A}b} = \eta_{ab} \,.
\end{align}
These Lorentz matrices encompass six degrees of freedom, representing three rotational degrees of freedom and three additional boost parameters. The three excess parameters are related to the ambiguity remaining in the choice of the worldline inside the compact body, which affects the definition of the angular momentum. This choice is to be fixed by imposing a spin supplementary condition (hereafter SSC) on the spin tensor, as explained in~\ref{sec:SSC}. Along the worldline, we also define in the usual way the antisymmetric rotation coefficients for the body-fixed tetrad 
\be
	\Omega^{\mu\nu} \equiv \epsilon^{A\mu}\frac{\uD \epsilon_{A}^{\ph{A}\nu}}{\ud \tau} \,,
\ee
where throughout the paper $\uD /\ud \tau \equiv u^{\nu}\nabla_{\nu}$.

From these definitions, we take as an Ansatz for the action of the point particle with spin that it depends kinematically only on $u^{\mu}$ and $\Omega^{\mu\nu}$, thus excluding from our formalism any dynamical departure from a solid body description (such as oscillations in neutron stars). The action will also depend on the metric around the position of the compact object. The implementation of finite-size effects is done by allowing a coupling to the derivatives of the metric as evaluated on the worldline. These derivatives can be written in a covariant form using the Riemann tensor and its symmetrized covariant derivatives~\cite{BI75} (in fact, only the Weyl tensor will contribute~\cite{DEF98}; in keeping with the literature, we write the formalism in terms of the Riemann tensor). Here, we are interested in the cubic-in-spins dynamics and we restrain ourselves to the octupolar order, i.e. to the first derivative of the Riemann tensor. We will describe in the appendix~\ref{app:generalization} the extension of this part of the formalism for general bodies with arbitrary derivative couplings. Our Ansatz for the action is therefore
\be\label{eq:deflagrangian}
	S = \int \ud \tau \, L\left[ u^{\mu}, \Omega^{\mu\nu}, g_{\mu\nu}, R_{\mu\nu\rho\sigma}, \nabla_{\lambda}R_{\mu\nu\rho\sigma} \right] \,.
\ee
The dynamical variables in the problem are the position $z^{\mu}$ and the tetrad $\epsilon_{A}^{\ph{A}\mu}$ (or equivalently the Lorentz matrices $\Lambda_{A}^{\ph{A}a}$). We define then the linear momentum $p_{\mu}$ and the spin tensor $S_{\mu\nu}$ as the conjuguate momenta of the positional and rotational degrees of freedom, and we also define the quadrupole moment $J^{\mu\nu\rho\sigma}$ and octupole moment $J^{\lambda\mu\nu\rho\sigma}$ according to
\begin{align}\label{eq:defpSJ}
	p_{\mu} &\equiv \frac{\partial L}{\partial u^{\mu}} \,, \quad S_{\mu\nu} \equiv 2 \frac{\partial L}{\partial \Omega^{\mu\nu}} \nn \\
	J^{\mu\nu\rho\sigma} &\equiv -6 \frac{\partial L}{\partial R_{\mu\nu\rho\sigma}} \,, \quad J^{\lambda\mu\nu\rho\sigma}\equiv -12 \frac{\partial L}{\partial \nabla_{\lambda}R_{\mu\nu\rho\sigma}} \,.
\end{align}
Notice that, from this very definition, $J^{\mu\nu\rho\sigma}$ has the same symmetries as the Riemann tensor, and $J^{\lambda\mu\nu\rho\sigma}$ is Riemann symmetric on its four last indices and also verifies, due to the Bianchi identity, $J^{[\lambda\mu\nu]\rho\sigma}=0$. These moments are non-dynamical, and the choice of names ``quadrupole moment'' and ``octupole moment'' will become clear from their leading order contribution to the metric around the body in section~\ref{sec:PNdynamics}. The relation of these multipole moments to Dixon's~\cite{Dixon64,Dixon73,Dixon74,Dixon79} is detailed in appendix~\ref{app:generalization}. It is crucial that the dynamical degrees of freedom enter the Lagrangian only through $u^{\mu}$ and $\Omega^{\mu\nu}$. The factors $-6$ and $-12$ are chosen for consistency of the notation with the literature (see the appendix~\ref{app:generalization}). Thus, the generic variation of $L$ will be
\be\label{eq:variationofLgeneric}
	\delta L = p_{\mu}\delta u^{\mu} + \frac{1}{2}S_{\mu\nu}\delta \Omega^{\mu\nu} + \frac{\partial L }{\partial g_{\mu\nu}} \delta g_{\mu\nu} - \frac{1}{6} J^{\mu\nu\rho\sigma} \delta R_{\mu\nu\rho\sigma} - \frac{1}{12} J^{\lambda\mu\nu\rho\sigma} \delta \nabla_{\lambda}R_{\mu\nu\rho\sigma}  \,.
\ee
%


\subsection{Scalar condition and homogeneity}

An important condition that we impose on this Lagrangian is that it must be a covariant scalar (the action will automatically be a Lorentz scalar with respect to transformations of the tetrad). As a consequence, by writing in~\eqref{eq:variationofLgeneric} the variation of all the tensors under an infinitesimal change of coordinates $x^{\mu} \rightarrow x^{\mu} + \zeta^{\mu}$, one obtains that the invariance of the Lagrangian imposes
\be\label{eq:scalarity}
	 2  \frac{\partial L}{\partial g_{\mu\nu}} = p^{\mu}u^{\nu} + S^{\mu\rho}\Omega^{\nu}_{\ph{\nu}\rho} + \frac{2}{3} R^{\mu}_{\ph{\mu}\lambda\rho\sigma}J^{\nu\lambda\rho\sigma} + \frac{1}{3} J^{\lambda\nu\tau\rho\sigma}\nabla_{\lambda}R^{\mu}_{\ph{\mu}\tau\rho\sigma} + \frac{1}{12} J^{\nu\lambda\tau\rho\sigma}\nabla^{\mu}R_{\lambda\tau\rho\sigma} \,.
\ee
This relation can be used to eliminate the partial derivative $\partial L/\partial g_{\mu\nu}$ from the variation in \eqref{eq:variationofLgeneric}.

The reparametrization invariance of the action~\eqref{eq:deflagrangian} with respect to the choice of $\tau$ has another consequence. The invariance of the action through a scaling $\tau \rightarrow \lambda \tau$ imposes that it must be homogeneous of degree one in $u^{\mu}$ and $\Omega^{\mu}$, which by Euler's theorem for homogeneous functions imposes
\be
	L = p_{\mu}u^{\mu}  + \frac{1}{2}S_{\mu\nu}\Omega^{\mu\nu} \,,
\ee
where all the dependence of the action on the Riemann tensor is hidden in the conjugate momenta $p_{\mu}$ and $S_{\mu\nu}$.


\subsection{Precession equation}

The precession equation, or evolution equation for the spins, is obtained by varying the action~\eqref{eq:deflagrangian} with respect to the internal freedom of the body-fixed tetrad, keeping fixed the worldline and the background tetrad $e_{a}^{\ph{a}\mu}$ (hence, the metric). This variation is represented by the antisymmetric quantities
\be
	\delta \theta^{ab} \equiv \Lambda^{Aa} \delta\Lambda_{A}^{\ph{A}b} \,,
\ee
which have six degrees of freedom, like the Lorentz transform relating $e_{a}^{\ph{a}\mu}$ and $\epsilon_{A}^{\ph{A}\mu}$. The variation of the components $\Omega^{ab} = e^{a}_{\ph{a}\mu}e^{b}_{\ph{b}\nu}\Omega^{\mu\nu}$ is given by
\be
	\delta \Omega^{ab} = e^{a}_{\ph{a}\mu}e^{b}_{\ph{b}\nu} \frac{\uD \delta\theta^{\mu\nu}}{\ud \tau} + \Omega^{a}_{\ph{a}c}\delta\theta^{cb} - \Omega^{b}_{\ph{b}c}\delta\theta^{ca}\,,
\ee
where we use $\delta\theta^{\mu\nu} = e_{a}^{\ph{a}\mu}e_{b}^{\ph{b}\nu}\delta\theta^{ab}$. By writing this variation in~\eqref{eq:variationofLgeneric}, doing the appropriate integrations by parts and using the fact that $\delta\theta^{ab}$ is arbitrary, one obtains
\be\label{eq:precessionOmega}
	\frac{\uD S^{\mu\nu}}{\ud \tau} = \Omega^{\mu}_{\ph{\mu}\rho}S^{\nu\rho} - \Omega^{\nu}_{\ph{\nu}\rho}S^{\mu\rho} \,.
\ee
Notice that this precession equation is universal in this form, in the sense that it depends only on the kinematic structure for the body-fixed tetrad in the Lagrangian. It is independent of the (possibly derivatives of) Riemann tensor couplings in the Lagrangian, and this remains valid at any multipolar order (see the appendix~\ref{app:generalization}). A direct consequence of~\eqref{eq:precessionOmega} is the existence of a conserved spin length, defined as
\be
	s^{2} \equiv \frac{1}{2} S_{\mu\nu}S^{\mu\nu} \,.
\ee
Its conservation $\ud s/\ud\tau = 0$ follows directly from $S^{\rho}_{\ph{\rho}\mu}\Omega^{\mu\nu}S_{\nu\rho} = 0$, and is very general as we have seen. In particular, it is also fully independent of the spin supplementary condition (see section~\ref{sec:SSC}).

The result~\eqref{eq:precessionOmega} can also be rewritten as the conservation of the body-frame tetrad components of the spin tensor $S^{AB} = \epsilon^{A}_{\ph{A}\mu}\epsilon^{B}_{\ph{B}\nu}S^{\mu\nu}$ (which are scalars for $\nabla$)
\be
	\frac{\ud S^{AB}}{\ud\tau} = 0 \,.
\ee
By taking into account the scalar condition~\eqref{eq:scalarity}, \eqref{eq:precessionOmega} can be rewritten in a form that now depends on the Riemann tensor couplings and where $\Omega^{\mu\nu}$ has been eliminated:
\begin{align}\label{eq:precessionScalarity}
	\frac{\uD S^{\mu\nu}}{\ud \tau} = 2p^{[\mu}u^{\nu]} + \frac{4}{3} R^{[\mu}_{\ph{\mu}\lambda\rho\sigma}J^{\nu]\lambda\rho\sigma} + \frac{2}{3}\nabla^{\lambda}R^{[\mu}_{\ph{\mu}\tau\rho\sigma} J_{\lambda}^{\ph{\lambda}\nu]\tau\rho\sigma} + \frac{1}{6}\nabla^{[\mu}R_{\lambda\tau\rho\sigma} J^{\nu]\lambda\tau\rho\sigma} \,.
\end{align}
This is the form of the spin precession equation that is produced by the stress-energy tensor approach, and the result is well known at quadrupolar order (see e.g.~\cite{SP10}). The generalization to higher multipolar orders can be obtained straightforwardly by adding the necessary terms in the scalar condition~\eqref{eq:scalarity} (see the appendix~\ref{app:generalization}).


\subsection{Equation of motion}

The equation of motion is obtained by varying with respect to the worldline of the particle~\cite{BI75}. To this end, we use the same setup as the one commonly used to derive the geodesic deviation equation. We consider a family of worldlines, defined in the vicinity of the original worldline, forming a 2-surface on which we introduce coordinates $(\tau, \lambda)$ and a vector $\xi^{\mu}$ such that $(u,\xi) = (\partial_{\tau},\partial_{\lambda})$ and $\calL_{u}\xi^{\mu} = 0$. We thus have the commutation relation $\xi^{\rho}\nabla_{\rho}u^{\mu}=u^{\rho}\nabla_{\rho}\xi^{\mu}$.

We can now use the requirement that the Lagrangian has to be a scalar to keep the variation entirely covariant, by writing
\be
	\delta{S} = \int \ud \tau \xi^{\rho}\partial_{\rho}L = \int \ud \tau \xi^{\rho}\nabla_{\rho}L \,.
\ee
The relevant variations (rewritten as $\delta \equiv \xi^{\rho} \nabla_{\rho}$) then take the simple forms
\begin{align}
	\delta u^{\mu} &= \frac{\uD \xi^{\mu}}{\ud \tau} \,, \nn \\
	\delta \Omega^{\mu\nu} &= - \frac{\uD}{\ud\tau}\left( \epsilon^{A\mu} \delta \epsilon_{A}^{\ph{A}\nu} \right) + \epsilon^{A\mu} \frac{\uD \delta \epsilon_{A}^{\ph{A}\nu}}{\ud\tau} - \epsilon^{A\nu} \frac{\uD \delta \epsilon_{A}^{\ph{A}\mu}}{\ud\tau} - \xi^{\rho}u^{\sigma}R^{\mu\nu}_{\ph{\mu\nu}\rho\sigma}  \,, \nn \\
	\delta R_{\mu\nu\rho\sigma} &= \xi^{\lambda}\nabla_{\lambda} R_{\mu\nu\rho\sigma} \,, \quad \delta \nabla_{\lambda}R_{\mu\nu\rho\sigma} = \xi^{\tau}\nabla_{\tau}\nabla_{\lambda} R_{\mu\nu\rho\sigma} \,,
\end{align}
and we have simply $\delta g_{\mu\nu} = 0$. Notice that the variation of the tetrad $\delta \epsilon_{A}^{\ph{A}\mu} = \xi^{\rho}\nabla_{\rho} \epsilon_{A}^{\ph{A}\mu}$ with the worldline is freely specifiable. One can for instance impose parallel transport from the original worldline to the perturbed one and assume $\delta \epsilon_{A}^{\ph{A}\mu}=0$. By writing these variations in~\eqref{eq:variationofLgeneric}, after integration by parts and using the arbitrariness of the displacement $\xi^{\mu}$, one obtains
\be\label{eq:resultEOM}
	\frac{\uD p_{\mu}}{\ud\tau} = -\frac{1}{2}R_{\mu\nu\rho\sigma}u^{\nu}S^{\rho\sigma} - \frac{1}{6} J^{\lambda\nu\rho\sigma} \nabla_{\mu} R_{\lambda\nu\rho\sigma} - \frac{1}{12} J^{\tau\lambda\nu\rho\sigma}\nabla_{\mu}\nabla_{\tau}R_{\lambda\nu\rho\sigma}\,,
\ee
a result well-known to the quadrupolar order (already present in~\cite{BI75}).
By keeping the variation of the tetrad unspecified, we would obtain, in factor of $\epsilon^{A\mu}\delta \epsilon_{A}^{\ph{A}\nu}$, a term which vanishes identically by taking into account~\eqref{eq:precessionOmega}. Again, the generalization to higher multipolar order is straightforward (see the appendix~\ref{app:generalization}).

 
\subsection{Stress-energy tensor}

We consider the 4-dimensional rewriting of the action~\eqref{eq:deflagrangian}, using a Dirac delta:
\be\label{eq:lagrangian4d}
	S = \int \ud^{4}x \sqrt{-g(x)} \int \ud \tau \, L\left[ u^{\mu}, \Omega^{\mu\nu}, g_{\mu\nu}, R_{\mu\nu\rho\sigma}, \nabla_{\lambda}R_{\mu\nu\rho\sigma} \right] \frac{\delta^{4}(x-z(\tau))}{\sqrt{-g(x)}} \,.
\ee
To obtain the stress-energy tensor of the particle, we keep fixed the worldline and the matrices $\Lambda_{A}^{\ph{A}a}$, and we vary with respect to the tetrad $e_{a}^{\ph{a}\mu}$. We define as usual the stress-energy tensor as the functional derivative
\be
	T^{\mu\nu} \equiv \frac{1}{\sqrt{-g}} e^{a(\mu}\frac{\delta S}{\delta e^{a}_{\ph{a}\nu)}} \,.
\ee
The individual variations are given by
\begin{align}
	\delta g_{\mu\nu} &= 2 e_{a(\mu}\delta e^{a}_{\ph{a}\nu)} \,, \nn \\
	\delta \Omega^{\mu\nu} &= -2\Omega^{\rho[\mu}\delta e_{a}^{\ph{a}\nu]} e^{a}_{\ph{a}\rho} + \frac{\uD}{\ud \tau}\left( e^{a[\mu}\delta e_{a}^{\ph{a}\nu]} \right) - u_{\sigma}\nabla^{[\mu}\left( \delta e_{a}^{\ph{a}\nu]} e^{a\sigma} \right) - u_{\sigma}\nabla^{[\mu}\left( e_{a}^{\ph{a}\nu]} \delta e^{a\sigma} \right) \,,
\end{align}
and we also have
\begin{align}\label{eq:variationRwithg}
	J^{\mu\nu\rho\sigma}\delta R_{\mu\nu\rho\sigma} &= J^{\mu\nu\rho\sigma}\left( 2 \nabla_{\mu}\nabla_{\sigma}\delta g_{\nu\rho} + R^{\lambda}_{\ph{\lambda}\nu\rho\sigma}\delta g_{\mu\lambda}\right) \,, \nn \\
	J^{\lambda\mu\nu\rho\sigma} \delta \nabla_{\lambda} R_{\mu\nu\rho\sigma} &= 2 J^{\lambda\rho\mu\nu\sigma}\nabla_{\lambda}\nabla_{\rho}\nabla_{\sigma}\delta g_{\mu\nu} + \delta g_{\mu\nu} J^{\lambda\mu\xi\rho\sigma} \nabla_{\lambda} R^{\nu}_{\ph{\nu}\xi\rho\sigma} \nn \\
	& \quad + \left(2 J^{\mu\nu\xi\rho\sigma} R^{\lambda}_{\ph{\lambda}\xi\rho\sigma} - 2 J^{\mu\lambda\xi\rho\sigma} R^{\nu}_{\ph{\nu}\xi\rho\sigma} - J^{\lambda\mu\xi\rho\sigma} R^{\nu}_{\ph{\nu}\xi\rho\sigma} \right) \nabla_{\lambda} \delta g_{\mu\nu} \,,
\end{align}
By writing the variation in~\eqref{eq:variationofLgeneric}, we get that the term proportional to $e^{a[\mu}\delta e_{a}^{\ph{a}\nu]}$ cancels again identically by virtue of the precession equation~\eqref{eq:precessionOmega}. We also use the scalar condition~\eqref{eq:scalarity} to replace the derivative $\partial L/\partial g_{\mu\nu}$. 

As a result of the variation, after integrations by parts one obtains the well-known pole-dipole contribution
\begin{align}\label{eq:Tmunupoledipole}
	T^{\mu\nu}_{\mathrm{pole-dipole}} &= \int \ud \tau p^{(\mu}u^{\nu)} \frac{\delta^{4}(x-z)}{\sqrt{-g}}   - \nabla_{\rho} \int \ud \tau S^{\rho(\mu}u^{\nu)} \frac{\delta^{4}(x-z)}{\sqrt{-g}} \,,
\end{align}
as well as the quadrupolar terms
\begin{align}\label{eq:Tmunuquad}
	T^{\mu\nu}_{\mathrm{quad}} &= \int \ud \tau \frac{1}{3} R^{(\mu}_{\ph{\mu}\lambda\rho\sigma}J^{\nu)\lambda\rho\sigma} \frac{\delta^{4}(x-z)}{\sqrt{-g}} - \nabla_{\rho}\nabla_{\sigma} \int \ud \tau \frac{2}{3} J^{\rho(\mu\nu)\sigma} \frac{\delta^{4}(x-z)}{\sqrt{-g}} \,,
\end{align}
which agrees with~\cite{SP10,BFH12} at quadrupolar order, and the octupolar contribution
\begin{align}\label{eq:Tmunuoct}
	T^{\mu\nu}_{\mathrm{oct}} &= \int \ud \tau \left[ \frac{1}{6} \nabla^{\lambda}R^{(\mu}_{\ph{\nu}\xi\rho\sigma}J_{\lambda}^{\ph{\lambda}\nu)\xi\rho\sigma} + \frac{1}{12} \nabla^{(\mu}R_{\xi\tau\rho\sigma}J^{\nu)\xi\tau\rho\sigma} \right] \frac{\delta^{4}(x-z)}{\sqrt{-g}} \nn \\
	& \quad  + \nabla_{\rho} \int \ud \tau \left[-\frac{1}{6} R^{(\mu}_{\ph{\mu}\xi\lambda\sigma}J^{|\rho|\nu)\xi\lambda\sigma} - \frac{1}{3}R^{(\mu}_{\ph{\mu}\xi\lambda\sigma}J^{\nu)\rho\xi\lambda\sigma} +\frac{1}{3} R^{\rho}_{\ph{\lambda}\xi\lambda\sigma}J^{(\mu\nu)\xi\lambda\sigma} \right] \frac{\delta^{4}(x-z)}{\sqrt{-g}} \nn \\
	& \quad + \nabla_{\lambda}\nabla_{\rho}\nabla_{\sigma} \int \ud \tau \frac{1}{3}J^{\sigma\rho(\mu\nu)\lambda}\frac{\delta^{4}(x-z)}{\sqrt{-g}} \,.
\end{align}
When we will apply this formalism to compact object binaries and compute the leading-order cubic-in-spin effects, only the last term with triple derivatives will matter, as the other ones will be subdominant in a post-Newtonian context. One can see this already from the occurence of a Riemann tensor, which introduces a relative 1PN order at least, whereas the two additional derivatives in the last term can be spatial derivatives which leave the PN order unaffected.


\section{Spin supplementary condition, definition of the mass and conserved norm spin vector}\label{sec:SSC}

The three extraneous degrees of freedom contained in the Lorentz matrices~\eqref{eq:defLambda} require to impose a spin supplementary condition (SSC) to close the system of equation describing the dynamics. This is already true in special relativity~\cite{HR74} and has been established early on~(see  e.g. \cite{CP51,Tulczyjew59}). The freedom in the choice of the SSC is related to the freedom in the choice of the worldline representing the motion inside the body. The SSC takes the form $V_{\nu}S^{\mu\nu} = 0$, with $V^{\mu}$ a timelike vector, which indeed imposes three additional constraints. A panorama of the different possible choices and their interpretation is given in~\cite{KS07}. In line with our previous work, we use in this article the Tulczyjew covariant spin supplementary condition~\cite{Tulczyjew59}
\be\label{eq:SSC}
	S^{\mu\nu}p_{\nu} = 0 \,.
\ee
The choice of SSC plays an important role in establishing a Hamiltonian formalism for a spinning particle~\cite{BRB09,Steinhoff11}, and care must be taken when working with a reduced Hamiltonian (or Routhian) about how it is enforced~\cite{HSS12}. In our case, however, since we are working at the level of the equations of motion and not at the level of a reduced Routhian or Hamiltonian, we will simply enforce it by hand. The use of different SSCs leads however to the spin variables to be a priori different, and this must be taken into account when comparing results obtained in different formalisms as we do in appendix~\ref{app:ADMEFT}. 

Before investigating the consequences of~\eqref{eq:SSC}, let us introduce a mass $m$ as the norm of the linear momentum, according to
\be
	m^{2} \equiv - p_{\mu}p^{\mu} \,.
\ee
Notice that this mass $m$ will not be conserved in general, as shown below. In the following, we will work perturbatively up to the cubic order in spin. At linear order in spin, we have the simple correspondence $p^{\mu} = m u^{\mu} + \calO(S^{2})$, hence also $S^{\mu\nu}u_{\nu} =  \calO(S^{3})$. Anticipating on the section~\ref{sec:spininduced}, we will also write that for the spin-induced quadrupole, $J^{\mu\nu\rho\sigma} = \calO(S^{2})$. 

From the conservation along the worldline of $S^{\mu\nu}p_{\nu} = 0$, using the equations of motion and precession~\eqref{eq:resultEOM} and~\eqref{eq:precessionScalarity}, we obtain
\begin{align}
	(u^{\nu}p_{\nu})p^{\mu} + m^{2} u^{\mu} &= \frac{1}{2} S^{\mu\nu}R_{\nu\lambda\rho\sigma}u^{\lambda}S^{\rho\sigma} + \frac{4}{3} p_{\nu} R^{[\nu}_{\ph{\nu}\lambda\rho\sigma}J^{\mu]\lambda\rho\sigma} \nn\\
	& \quad + \frac{1}{6}S^{\mu\nu}J^{\lambda\tau\rho\sigma}\nabla_{\nu}R_{\lambda\tau\rho\sigma} + \frac{1}{12}S^{\mu\nu}J^{\xi\lambda\tau\rho\sigma}\nabla_{\nu}\nabla_{\xi}R_{\lambda\tau\rho\sigma} \nn\\
	& \quad + \frac{2}{3} p_{\nu} \nabla^{\lambda}R^{[\nu}_{\ph{[\nu}\tau\rho\sigma}J_{\lambda}^{\ph{\lambda}\mu]\tau\rho\sigma} + \frac{1}{6} p_{\nu} \nabla^{[\nu}R_{\lambda\tau\rho\sigma}J^{\mu]\lambda\tau\rho\sigma} + \calO(S^{4}) \,.
\end{align}
Contracted again with $u_{\mu}$, this gives $p_{\mu}u^{\mu} = -m + \calO(S^{4})$, and from there the relation between $p^{\mu}$ and $u^{\mu}$
\begin{align}\label{eq:relationpu}
	p^{\mu} &= m u^{\mu} - \frac{1}{2m} S^{\mu\nu}R_{\nu\lambda\rho\sigma}u^{\lambda}S^{\rho\sigma} + \frac{4}{3} R^{[\mu}_{\ph{\mu}\lambda\rho\sigma}J^{\nu]\lambda\rho\sigma} u_{\nu} \nn\\
	& \quad + \frac{2}{3} u_{\nu} \nabla^{\lambda}R^{[\mu}_{\ph{[\nu}\tau\rho\sigma}J_{\lambda}^{\ph{\lambda}\nu]\tau\rho\sigma} + \frac{1}{6} u_{\nu} \nabla^{[\mu}R_{\lambda\tau\rho\sigma}J^{\nu]\lambda\tau\rho\sigma} - \frac{1}{6m}S^{\mu\nu}J^{\lambda\tau\rho\sigma}\nabla_{\nu}R_{\lambda\tau\rho\sigma} + \calO(S^{4}) \,.
\end{align}
We will see in section~\ref{sec:PNdynamics} that, at the leading 3.5PN order for the cubic in spin effects, the additional terms in this relation do contribute. 

We can now turn to the equation of evolution of the mass $m$ itself. By using~\eqref{eq:resultEOM} and~\eqref{eq:relationpu}, we get 
\be\label{eq:dmdtau}
	\frac{\ud m}{\ud \tau} = \frac{1}{2m} p^{\mu}u^{\nu}R_{\mu\nu\rho\sigma}S^{\rho\sigma} + \frac{1}{6} J^{\lambda\nu\rho\sigma}  \frac{\uD}{\ud \tau}R_{\lambda\nu\rho\sigma} + \frac{1}{12} J^{\tau\lambda\nu\rho\sigma}  \frac{\uD }{\ud \tau}\nabla_{\tau}R_{\lambda\nu\rho\sigma} +\calO(S^{4}) \,.
\ee
At quadratic order in spin, under the hypothesis that $\uD J^{\mu\nu\rho\sigma}/\ud \tau = \calO(S^{3})$, which will be verified if $J^{\mu\nu\rho\sigma}$ can be expressed with $S^{\mu\nu}$, $u^{\mu}$ only and is quadratic in spin (as for the spin-induced quadrupole~\eqref{eq:defquadrupole} below), one can define a conserved mass according to
\be\label{eq:defconservedmass}
	\tilde{m} \equiv - p_{\mu}u^{\mu} - \frac{1}{6} J^{\lambda\nu\rho\sigma} R_{\lambda\nu\rho\sigma} \,, \quad \frac{\ud \tilde{m}}{\ud \tau} = \calO(S^{3}) \,.
\ee
The additional term in this conserved mass is found to contribute at 3PN, and must be taken into account at this order~\cite{BFMP15}. However, since we will be working at leading PN order for the SSS effects, this contribution can be ignored as it is of a relative 1PN order.

At cubic order in spin, by contrast, one can see that the two first terms of~\eqref{eq:dmdtau} yield both $\calO(S^{3})$ contributions that cannot be rewritten as a single time derivative, therefore it does not seem possible to absorb the right-hand side in a redefinition of the mass. However, we checked that these additional SSS terms intervene only at the 4.5PN order, that is at 1PN relative order, and we can safely ignore them in our leading-order study.

Notice that here we keep the form of the Lagrangian unspecified; if one were to write explicitly a particular Lagrangian~\eqref{eq:deflagrangian}~\cite{SP12}, the relation between $p^{\mu}$ and $u^{mu}$ would follow, and a conserved mass could appear simply as a constant parameter in the action.

After having specified the SSC~\eqref{eq:SSC}, one can also construct a spin vector $S_{\mu}$, using the fact that it is normal to the unit vector $p^{\mu}/m$, according to
\be
	S_{\mu} = - \frac{1}{2} \epsilon_{\mu\nu\rho\sigma}\frac{p^{\nu}}{m}S^{\rho\sigma} \,, \quad S^{\mu\nu} = \epsilon^{\mu\nu\rho\sigma}\frac{p_{\rho}}{m}S_{\sigma} \,,
\ee
with $\epsilon_{\mu\nu\rho\sigma}$ the Levi-Civita tensor\footnote{We adopt the convention $\epsilon_{\mu\nu\rho\sigma} = \sqrt{-g}[\mu\nu\rho\sigma]$ with the antisymmetric symbol $[0123] = +1$. Notice that this differs from the convention of~\cite{BMFB13}.}. Using $p_{\mu}S^{\mu} = 0$ and $S_{\mu}S^{\mu} = s^{2}$, one can build a Euclidean conserved norm spin vector that will be very convenient for the presentation of PN results, following a geometric prescription described in~\cite{BMFB13} (see also~\cite{DJSspin}). An orthonormal tetrad $e_{A} = (e_{0},e_{I})$ is introduced, with $e_{0}^{\ph{0}\mu} = p^{\mu}/m$, so that $s^{2} = \delta^{IJ}S_{I}S_{J}$ with $S_{I} = e_{I}^{\ph{I}\mu}S_{\mu}$. The Euclidean conserved norm spin vector is then identified to $S_{I}$, with the prescription that $e_{Ii}$ is the symmetric square root of $\gamma_{ij} = g_{ij} + p_{i}p_{j}/m^{2}$. This spin vector then verifies a precession equation of the form $\dot{\bm{S}} = \bm{\Omega}\times \bm{S}$~\cite{BMFB13,BFMP15}.

In practice, it turns out that the corrections in the relation between the spin tensor and this spin vector are of order $\calO(2)$ for the SO terms, $\calO(5)$ for the SS terms and $\calO(6)$ for the SSS terms. Since we will be working with a leading-order scheme, at $\calO(0)$ for the NS terms, $\calO(3)$ for the SO terms, $\calO(4)$ for the SS terms, and finally $\calO(7)$ for the SSS terms, we will not need to compute these contributions. We will therefore use only the leading order definition of this conserved norm spin vector, which is simply (with $i,j,k$ spatial Euclidean indices and $S^{ij}$ the spatial components of the spin tensor $S^{\mu\nu}$, see section~\ref{sec:PNdynamics})
\be
	S_{i} = \frac{1}{2} \varepsilon_{ijk}S^{jk} \,, \quad S^{ij} = \varepsilon^{ijk}S^{k}\,.
\ee
%


\section{Spin-induced multipoles}\label{sec:spininduced}

We now investigate the structure of the quadrupole and octupole moments if they are induced by the rotation of the object only. Thus, our central assumption will be that these multipoles can be expressed only with the spin tensor, mass and velocity (or linear momentum). The leading order couplings in the Lagrangian have been written down up to the quartic-in-spin order in~\cite{LS14b}. Note that additionally to such spin-induced effects, tidal effects can be represented in an effective action formalism as well~\cite{BDF12}.

An important point is that we can in fact remove all dependency of the action on the Ricci tensor $R_{\mu\nu}$, so that only the Weyl tensor $C_{\mu\nu\rho\sigma}$ intervenes instead of the Riemann tensor. This has been stressed in~\cite{DEF98} (see also~\cite{GR06} in the EFT context) and is due to the fact that a quantity that is proportional to the field equations can be absorbed in a local redefinition of the fields and dynamical variables.

We will work in a formal expansion in spin, and our goal is to write a census of the possible scalars in the action that can be built at each order in spin, allowing us to deduce the associated multipole moments. We shall write the couplings in the Lagrangian directly in terms of the linear momentum $p^{\mu}$ and spin tensor $S^{\mu\nu}$, which were defined as the conjugate momenta of $u^{\mu}$ and $\Omega^{\mu\nu}$ in~\ref{sec:formalism}. We can for instance assume that the action takes the form~\cite{HR74,Porto06,Steinhoff11,BFH12}
\be
	L = -m u + \frac{I}{4u} \Omega_{\mu\nu}\Omega^{\mu\nu} + \dots \,,
\ee
with $I$ the inertia momentum and where $u = \sqrt{-u_{\mu}u^{\mu}}$ ensures the homogeneity of the action. From here we can invert order by order to replace $u^{\mu}$ and $\Omega^{\mu\nu}$ by
\be
	u^{\mu} = p^{\mu}/m + \dots \,, \quad \Omega^{\mu\nu} = S^{\mu\nu}/I + \dots \,.
\ee
Here the ellipses do not represent an expansion in the spin, but rather a perturbative PN expansion.

Since the multipoles are given by derivatives of the Lagrangian with respect to the Riemann tensor, terms in the action displaying the contraction $V^{\mu} = p_{\nu}S^{\mu\nu}$ will give contributions in the multipoles that will vanish after imposing the SSC $V^{\mu} = 0$ at the level of the equations of motion\footnote{See~\cite{Levi08,Steinhoff11,HSS12} for the imposition of the SSC at the level of the action.}. We therefore ignore this possible terms and use the orthogonality relation of the SSC to reduce the possible choices for the action. Since $p^{\mu} = m u^{\mu} + \calO(S^{2})$, we will use $u^{\mu}$ instead of the unit vector $p^{\mu}/m$ for convenience\footnote{The distinction should be made, however, at quartic order in spin.}.

So far, we have left still considerable freedom in the Lagrangian introduced in section~\ref{sec:formalism}, but we will also assume that dimensionful quantities (such as the inertia momentum $I$) appear only in the expected way, such that only the spin, the mass and the velocity (or $p/m$) appear in the expansion in spin of the action. With these restrictions, dimensional analysis shows readily that the structure at quadrupolar order will be $RSS/m$, and at octupolar order $\nabla R SSS/m^{2}$. Notice that at quartic order, two terms can appear, $RRSSSS/m^{3}$ (contributing to the quadrupole) and $\nabla\nabla R SSSS/m^{3}$ (contributing to the hexadecapole). Also, one should remember that there is in general more than one definition of the mass, as shown in section~\ref{sec:SSC}.


\subsection{Quadrupolar order}

The form of the spin-induced quadrupole is already well known, but we present here a derivation for completeness. It will be useful to introduce the Bel decomposition of the Weyl tensor~\cite{Bel58} with respect to $u^{\mu}$. One defines the electric and magnetic components of the Weyl tensor as\footnote{Various conventions exist in the literature; here, we follow those of~\cite{BDF12,BFH12}.}
\begin{align}\label{eq:Bel}
	G_{\mu\nu} &\equiv - C_{\mu\alpha\nu\beta}u^{\alpha}u^{\beta} \,, \nn\\
	H_{\mu\nu} &\equiv 2 {}^{*}C_{\mu\alpha\nu\beta}u^{\alpha}u^{\beta} \,,
\end{align}
with ${}^{*}C_{\mu\nu\rho\sigma} \equiv 1/2 \epsilon_{\mu\nu\alpha\beta}C^{\alpha\beta}_{\ph{\alpha\beta}\rho\sigma}$ the dual of the Weyl tensor. They are both symmetric, traceless and orthogonal to $u^{\mu}$, and they have thus 5 independent components each, which matches the 10 components of the Weyl tensor. One can build an inverse formula, expressing the Weyl tensor with these tidal tensors as
\begin{align}\label{eq:Belinverse}
	C_{\mu\nu\rho\sigma} &= 4 g_{\gamma[\mu}G_{\nu][\rho}\delta_{\sigma]}^{\ph{\sigma]}\gamma} + 8 u_{[\mu}G_{\nu][\rho}u_{\sigma]} \nn\\
	&+ \epsilon_{\mu\nu}^{\ph{\mu\nu}\lambda\tau}u_{\lambda}H_{\tau[\rho}u_{\sigma]} +  \epsilon_{\rho\sigma}^{\ph{\rho\sigma}\lambda\tau}u_{\lambda}H_{\tau[\mu}u_{\nu]} \,,
\end{align}
which is equivalent, by replacing the tidal tensors, to
\be\label{eq:decRiemann}
	C_{\mu\nu\rho\sigma} = 4 g_{\gamma[\mu}C_{\nu]\alpha\beta[\rho}\delta_{\sigma]}^{\ph{\sigma]}\gamma}u^{\alpha}u^{\beta} +  2 C_{\mu\nu\alpha[\rho}u_{\sigma]}u^{\alpha} + 2C_{\rho\sigma\alpha[\mu}u_{\nu]}u^{\alpha} \,.
\ee
Notice that this decomposition is dimension-dependent. In the following, we will use the following convenient notation for the square of the spin tensor
\be
	\Theta^{\mu\nu}\equiv S^{\mu\lambda}S^{\nu}_{\ph{\nu}\lambda} \,,
\ee
which is symmetric.

Because $u^{\mu}$ is orthogonal to $S^{\mu\nu}$, at first glance we have only three possible couplings (excluding parity-violating couplings featuring $\epsilon_{\mu\nu\rho\sigma}$), given by~\cite{Porto06}
\begin{align}
	(i) & \; C_{\mu\nu\rho\sigma} S^{\mu\rho}u^{\nu}u^{\sigma} \nn\\
	(ii) & \; C_{\mu\nu\rho\sigma} S^{\mu\nu}S^{\rho\sigma}\nn\\
	(iii) & \; C_{\mu\nu\rho\sigma} S^{\mu\rho}S^{\nu\sigma} \,. \nn
\end{align}
The terms $(ii)$ and $(iii)$ are in fact related by the Bianchi identity, $(ii) = 2(iii)$, and the decomposition~\eqref{eq:decRiemann} readily shows that they are amenable to the first term $(i)$.

Thus, we are left with a single coupling in the Lagrangian, with an undetermined constant that we denote by $\kappa$ ($C_{\rm{ES}^{2}}$ in~\cite{Porto06,LS14b}), introducing $u$ for homogeneity,
\be
	\left( L \right)_{C} = \frac{\kappa}{2 m u} C_{\mu\nu\rho\sigma} u^{\mu}S^{\nu}_{\ph{\nu}\xi}S^{\rho\xi}u^{\sigma} = \frac{\kappa}{2 m}G_{\mu\nu}\Theta^{\mu\nu} \,,
\ee
The resulting expression for the spin-induced quadrupolar moment $J^{\mu\nu\rho\sigma} = -6\;\partial L /\partial R_{\mu\nu\rho\sigma}$ is
\be\label{eq:defquadrupole}
	J^{\mu\nu\rho\sigma} = \frac{3 \kappa}{m}u^{[\mu}S^{\nu]\lambda}S_{\lambda}^{\ph{\lambda}[\rho}u^{\sigma]} \,,
\ee
for which one can check that it has the symmetries of the Riemann tensor and also satisfies the Bianchi identity $J^{[\mu\nu\rho]\sigma} = 0$.


\subsection{Octupolar order}

At the octupolar order, one can introduce a generalization of the previous decomposition according to
\begin{align}
	G_{\mu\nu\rho} &\equiv - \nabla^{\perp}_{(\mu} C_{\nu|\alpha|\rho)\beta}u^{\alpha}u^{\beta} \,, \nn\\
	H_{\mu\nu\rho} &\equiv 2 \nabla^{\perp}_{(\mu} {}^{*}C_{\nu|\alpha|\rho)\beta}u^{\alpha}u^{\beta} \,,
\end{align}
with $\nabla^{\perp}_{\mu} = (\delta_{\mu}^{\ph{\mu}\nu}+u_{\mu}u^{\nu})\nabla_{\nu}$ the derivative projected othogonally to $u^{\mu}$. The tensors $G_{\mu\nu\rho}$ and $H_{\mu\nu\rho}$ are again symmetric, trace-free and orthogonal to $u^{\mu}$. It is possible to write down decomposition formulas analogous to~\eqref{eq:Belinverse} and~\eqref{eq:decRiemann}, using also the projections, similar to~\eqref{eq:Bel}, of the derivative $\uD C_{\mu\nu\rho\sigma} /\ud\tau$ (see for instance~\cite{PV10} for expressions in a particular coordinate system). However, we find that looking directly at the Weyl tensor will be simpler.

Terms with $\nabla_{\mu}C^{\mu}_{\ph{\mu}\nu\rho\sigma}$ are not allowed due to the vanishing of the Ricci tensor and the Bianchi identity. Now, from the parity of the number of indices, we see that there should be an odd number of $u$ contracted with $\nabla C$. With 3 $u$'s, one has $\uD/\ud \tau C_{\mu\alpha\nu\beta}u^{\alpha}u^{\beta}$, which is symmetric. But we can only contract it with $(S^{\rho\sigma}S_{\rho\sigma})S^{\mu\nu}$ and $S^{\mu\rho}S_{\rho\sigma}S^{\sigma\nu}$, both antisymmetric; among the terms with one contraction with $u$, similarly, we cannot have $\uD/\ud \tau R_{\mu\nu\rho\sigma}$ contracted with $S,\Theta$. We are left with three possible terms
\begin{align}
	(i) & \; \Theta^{\mu\nu}S^{\rho\sigma} \nabla_{\mu}C_{\nu\rho\sigma\alpha}u^{\alpha} \nn\\
	(ii) & \; \Theta^{\mu\sigma}S^{\nu\rho} \nabla_{\mu}C_{\nu\rho\sigma\alpha}u^{\alpha} \nn\\
	(iii) & \; \Theta^{\rho\sigma}S^{\mu\nu} \nabla_{\mu}C_{\nu\rho\sigma\alpha}u^{\alpha} \,,
\end{align}
which can be rewritten thanks to Bianchi identities to obtain $(i)=-1/2(ii)=(iii)$. We have thus only one possible coupling, which can be written in either one of the three forms above (introducing the constant $\lambda$, noted $C_{\rm{BS}^{3}}$ in~\cite{LS14b})
\be
	\left( L \right)_{\nabla C} = -\frac{\lambda}{12 m^{2}u^{2}} \nabla_{\lambda}C_{\mu\nu\rho\alpha} \Theta^{\lambda\rho}S^{\mu\nu}u^{\alpha} = -\frac{\lambda}{24 m^{2} u^{2}} H_{\mu\nu\rho}\epsilon^{\mu}_{\ph{\mu}\alpha\beta\gamma}\Theta^{\nu\rho}S^{\alpha\beta}u^{\gamma}  \,,
\ee
This expression is in agreement with the one proposed recently in~\cite{LS14b}\footnote{Notice a change of sign in the second equality due to the opposite signature of the metric in~\cite{LS14b}}.

Next, we evaluate the derivative $J^{\lambda\mu\nu\rho\sigma} = - 12 \;\partial{L}/{\partial\nabla_{\lambda}C_{\mu\nu\rho\sigma}}$ by finding a combination of the three different terms above that has all the required symmetries, including the two Bianchi identities. It is given by
\begin{align}\label{eq:defoctupole}
	J^{\lambda\mu\nu\rho\sigma} &= \frac{\lambda}{4 m^{2}} \left[ \Theta^{\lambda[\mu}u^{\nu]}S^{\rho\sigma} + \Theta^{\lambda[\rho}u^{\sigma]}S^{\mu\nu} \right.\nn\\
	& \qquad \quad \; - \Theta^{\lambda[\mu}S^{\nu][\rho}u^{\sigma]} - \Theta^{\lambda[\rho}S^{\sigma][\mu}u^{\nu]} \nn\\
	&\qquad \quad \; \left. - S^{\lambda[\mu}\Theta^{\nu][\rho}u^{\sigma]} - S^{\lambda[\rho}\Theta^{\sigma][\mu}u^{\nu]} \right] \,.
\end{align}

The constants $\kappa,\lambda$ characterize the deformation of the compact object due to its spin. In the black hole case, they can be fixed by comparison with an isolated Kerr black hole, for instance by comparing the leading multipoles of the PN metric to their Kerr expressions, or by comparing the resulting dynamics with the known orbits of a test particle in the Kerr background. Numerical factors have been arranged so that $\kappa=\lambda=1$ for black holes. For neutron stars however, a numerical computation for an isolated star is required~\cite{LP97}, yielding $\kappa\sim 4-8$ depending on the equation of state. The value of $\lambda$ for neutron stars is yet unknown.


\section{Leading order cubic-in-spin effects in the post-Newtonian dynamics}\label{sec:PNdynamics}


\subsection{General setting and definitions}

In this section and the next one, we will apply the formalism presented above to the computation of cubic-in-spin effects for inspiralling compact binaries, both in the dynamics and in the gravitational waves emitted. Both objects will be modelled as compact, point-like objects carrying a spin and a multipolar structure that is assumed to be induced by rotation only, as described above.

As in previous works~\cite{MBFB13,BMFB13,BMB13,MBBB13}, we use an approach in harmonic coordinates (see~\cite{Bliving} for a review). We work directly at the level of the equations of motion, by contrast with the other methods such as the ADM and EFT approaches, having in mind the application of these results to the multipolar post-Newtonian wave generation formalism in harmonic coordinates, which is done is the next section. To deal with the singularities introduced by the treatment of the two bodies as point particles, since we are working at leading order we can simply apply the Hadamard regularization procedure described in~\cite{BF00reg}.

The qualitative structure of the equations of motion, with various spin orders entering at various PN orders, is as follows (see~\cite{Bliving} for more details)
\be\label{eq:PNstructA}
	\bm{A} = \bm{A}_{\rm NS} + \frac{1}{c^{3}} \bm{A}_{\rm SO} + \frac{1}{c^{4}} \bm{A}_{\rm SS} + \frac{1}{c^{7}} \bm{A}_{\rm SSS} + \calO(8) \,,
\ee
where we only indicate the leading order terms (each spin order also has 1PN relative and higher corrections, and radiation reaction also enters at the 2.5PN relative order for each spin sector). The fact that the SSS contributions enter at the 3.5PN order, and not at the 2.5PN order, is natural when considering the structure of the Kerr metric, where cubic-in-spin terms arise as products of quadratic (2PN) and linear (1.5PN) terms. The conserved energy $E$ and the energy loss rate $\calF$ have the same structure as~\eqref{eq:PNstructA} for the PN order of the spin terms. Similarly, the structure of the precession equation for the conserved norm spin vectors is, writing $\dot{\bm{S}} = \bm{\Omega} \times \bm{S}$,
\be\label{eq:PNstructOmega}
	\bm{\Omega} = \frac{1}{c^{2}}\bm{\Omega}_{\rm NS} + \frac{1}{c^{3}}\bm{\Omega}_{\rm SO} + \frac{1}{c^{6}}\bm{\Omega}_{\rm SS} + \calO(7) \,. 
\ee
Now, since we are working at the leading order for the SSS terms in both the dynamics and energy flux, we see that we can in fact ignore all but the leading order terms in the NS, SO and SS sectors, and that only the SO $\calO(3)$ terms in $\bm{\Omega}$ must be taken into account\footnote{The only computation where the SS $\calO(6)$ would intervene would be the determination of a conserved angular momentum at $\calO(7)$, since its PN expansion starts as $\bm{J} = \bm{L} + \bm{S}/c$, with spin effects at the 0.5PN order instead of the usual 1.5PN order.}. These leading order NS, SO and SS terms do yield contributions at the SSS $\calO(7)$ order when a time derivative is taken (when building conserved quantities, or when applying the waveform generation formalism), upon replacing the equations of motion and precession. Thus, in the following we will include them in our presentation for consistency, although they are already known. In practice, working at leading order will greatly simplify many aspects of the formalism, as for instance the complication of the iteration of the solution of the field equations mainly depends on the relative order.

To describe the two bodies, we will denote by $m_{1}$, $m_{2}$ their masses (according to the discussion in~\ref{sec:SSC}, for our purpose we can consider them as conserved without further redefinition), by $y_{1}^{i}$, $y_{2}^{i}$ their positions, by $v_{1}^{i}$, $v_{2}^{i}$ their velocities, and by $S_{1}^{ij}$, $S_{2}^{ij}$ the spatial components of their spin tensors. We also use the notations $r_{12} = |\bfy_{1} - \bfy_{2}|$, $n_{12}^{i} = (y_{1}^{i} - y_{2}^{i})/|\bfy_{1} - \bfy_{2}|$, $v_{12}^{i}  = v_{1}^{i} - v_{2}^{i}$ and $\partial^{1}_{i} = \partial / \partial y_{1}^{i}$, where boldface denotes a vector. The operation $\exch$ means the exchange of the label of the two bodies. All spatial indices are raised and lowered using the background Euclidean metric $\delta_{ij}$, and we make no distinction between upper and lower indices. We also restore the factors $c$ in all expressions.

The covariant spin supplementary condition $p_{\nu}S^{\mu\nu} = 0$ translates at leading order to
\be\label{eq:SSCleading}
	S^{0i} = -S^{ij}\frac{v^{j}}{c} + \calO(3) \,,
\ee
and one can check that there are neither SS $\calO(3)$ nor SSS $\calO(4)$ contributions, so that we can simply use this leading order relation for our purposes. As discussed in section~\ref{sec:SSC}, we use conserved norm Euclidean spin vectors $S_{1}^{i}$, $S_{2}^{i}$, and the leading order relation to the spin tensor, given by $S^{ij} = \varepsilon^{ijk}S^{k}$, $S^{i} = \varepsilon^{ijk}S^{jk}/2$, will be sufficient.


\subsection{The metric potentials}

In harmonic coordinates, defined by the gauge condition $\partial_{\nu}h^{\mu\nu} = 0$ with $h^{\mu\nu} = \sqrt{-g}g^{\mu\nu} - \eta^{\mu\nu}$ the metric perturbation, Einstein equations become
\be\label{eq:Einsteinharmonic}
	\Box h^{\mu\nu} = \frac{16\pi G}{c^{4}} |g| T^{\mu\nu} + \Lambda^{\mu\nu}\left[ h \right] \equiv \frac{16\pi G}{c^{4}} \tau^{\mu\nu} \,,
\ee
with $\Box = \partial_{\mu}\partial^{\mu}$ the flat d'Alembert operator and $\Lambda^{\mu\nu}$ a non-compact support source that contains non-linearities in $h$. In the general case, the iteration of this equation yields a hierarchy of potentials, each obeying d'Alembert equations with sources of increasing complexity~\cite{BF01eom}. In our case, as we work at the leading post-Newtonian order of the spin contributions, we will only need the simplest of these potentials, namely
\begin{subequations}\label{eq:defVVi}
\begin{align}
V & = \Box_{\mathcal{R}}^{-1}[-4 \pi G\, \sigma]\;,\label{V}\\ 
V_{i} &= \Box_{\mathcal{R}}^{-1}[-4 \pi G\, \sigma_{i}]\,,
\end{align}
\end{subequations}
with $\Box_{\mathcal{R}}^{-1}$ the usual retarded inverse d'Alembertian and the following definitions for the sources (matter-only for these simplest potentials)
\be
	\sigma=\frac{1}{c^{2}}(T^{00}+T^{ii}) \,, \quad \sigma_{i}=\frac{1}{c}T^{0i} \,.
\ee
These metric potentials enter the metric components according to
\begin{subequations}\label{eq:metricg}
\begin{align} g_{00}
 &=  -1 + \frac{2}{c^{2}}V - \frac{2}{c^{4}} V^{2} + \calO(6) \,, \\ 
g_{0i} & = - \frac{4}{c^{3}}V_{i} + \calO(5) \,, \\ 
g_{ij} & = \delta_{ij} \left[1 + \frac{2}{c^{2}}V \right] + \calO(4) \,.
\end{align}
\end{subequations}
From the expressions of the spin-induced moments~\eqref{eq:defquadrupole}, \eqref{eq:defoctupole} and of the stress-energy tensor~\eqref{eq:Tmunupoledipole}, \eqref{eq:Tmunuquad} and~\eqref{eq:Tmunuoct}, we see that at each spin order only the term with the largest number of spatial derivatives contributes at leading PN order, yielding (with $\delta_{1,2} = \delta^{3}(\bfx -\bfy_{1,2})$)
\begin{subequations}\label{eq:sigmaLO}
\begin{align}
	\sigma^{\rm NS}&= m_{1} \delta_{1} + \exch + \calO(2) \,, \\
	\sigma^{\rm SO}&= -\frac{2}{c^{3}}S_{1}^{ij}v_{1}^{j}\partial_{i}\delta_{1} + \exch + \calO(5) \,, \\
	\sigma^{\rm SS}&= \frac{\kappa_{1}}{2 m_{1}c^{4}}S_{1}^{ik}S_{1}^{jk}\partial_{ij}\delta_{1} + \exch + \calO(6) \,, \\
	\sigma^{\rm SSS}&= \frac{\lambda_{1}}{3 m_{1}^{2}c^{7}}v_{1}^{m}S_{1}^{mi}S_{1}^{jl}S_{1}^{kl}\partial_{ijk}\delta_{1} + \exch + \calO(9) \,,
\end{align}
\end{subequations}
and
\begin{subequations}\label{eq:sigmaiLO}
\begin{align}
	\sigma^{\rm SO}_{i} &= \frac{1}{2c}S_{1}^{ij}\partial_{j}\delta_{1} + \exch + \calO(3) \,, \\
	\sigma^{\rm SSS}_{i} &= \frac{\lambda_{1}}{12 m_{1}^{2}c^{5}}S_{1}^{ij}S_{1}^{km}S_{1}^{lm}\partial_{jkl}\delta_{1} + \exch + \calO(7) \,.
\end{align}
\end{subequations}
For the odd orders in spin, the source $\sigma_{i}$ is at 1PN lower order than the source $\sigma$, which pushes the $V_{i}$ contributions at the same level as the $V$ contributions in all subsequent equations, whereas it only intervenes at 1PN relative order and can be neglected in the non-spinning and quadratic terms. This structure will also appear in the computation of source moments in section~\ref{subsec:sourcemoments}. Notice also that the SSC~\eqref{eq:SSCleading} was used in the leading-order $\sigma$ for these odd orders in spin. These sources give directly for the leading order potentials 
\begin{subequations}
\begin{align}
	V^{\rm NS}&= \frac{G m_{1}}{r_{1}} + \exch + \calO(2) \,, \\
	V^{\rm SO}&= -\frac{2 G}{c^{3}}S_{1}^{ij}v_{1}^{j}\partial_{i}\left( \frac{1}{r_{1}} \right) + \exch + \calO(5) \,, \\
	V^{\rm SS}&= \frac{G \kappa_{1}}{2 m_{1}c^{4}}S_{1}^{ik}S_{1}^{jk}\partial_{ij}\left( \frac{1}{r_{1}} \right) + \exch + \calO(6) \,, \\
	V^{\rm SSS}&= \frac{G \lambda_{1}}{3 m_{1}^{2}c^{7}}v_{1}^{m}S_{1}^{mi}S_{1}^{jl}S_{1}^{kl}\partial_{ijk}\left( \frac{1}{r_{1}} \right) + \exch + \calO(9) \,,
\end{align}
\end{subequations}
and
\begin{subequations}
\begin{align}
	V^{\rm SO}_{i} &= \frac{G}{2c}S_{1}^{ij}\partial_{j}\left( \frac{1}{r_{1}} \right) + \exch + \calO(3) \,, \\
	V^{\rm SSS}_{i} &= \frac{G \lambda_{1}}{12 m_{1}^{2}c^{5}}S_{1}^{ij}S_{1}^{km}S_{1}^{lm}\partial_{jkl}\left( \frac{1}{r_{1}} \right) + \exch + \calO(7) \,.
\end{align}
\end{subequations}
These results give the leading order quadrupolar and octupolar piece of the metric around each spinning object. If we compare this contribution to the metric to the multipoles parametrizing the Kerr metric (for instance as given in~\cite{Thorne80}), we can read that for Kerr black holes, $\kappa=\lambda=1$.


\subsection{Results for the dynamics}

We can now turn to the computation of the equation of motion at cubic order in spin and at the leading 3.5PN order. From the covariant expression~\eqref{eq:resultEOM}, we see that various contributions enter,
\be
	a_{1}^{i} \equiv \frac{\ud v_{1}^{i}}{\ud t} = \left( a_{1}^{i} \right)^{\mathrm{geod}} + \left( a_{1}^{i} \right)^{\mathrm{Papapetrou}} + \left( a_{1}^{i} \right)^{\mathrm{quad}} + \left( a_{1}^{i} \right)^{\mathrm{oct}} + \left( a_{1}^{i} \right)^{p \leftrightarrow u} \,.
\ee
Here the first term arises from the replacement in the metric in the Christoffel symbols generated by the covariant derivative, as in the case of geodesic motion; the second corresponds to the pole-dipole term, the first term in~\eqref{eq:resultEOM} that is known since the work of Mathisson and Papapetrou~\cite{Mathisson37,Mathisson40,Papapetrou51spin}; the third and the fourth to respectively the additional quadrupolar and octupolar terms in~\eqref{eq:resultEOM}; and the fifth to the contribution of the additional terms in the relation $p^{\mu}=m u^{\mu}+\dots$ given in~\eqref{eq:relationpu}. By gathering all these contributions, we obtain
\begin{align}\label{eq:a1iScube}
	\left( a_{1}^{i} \right)_{\rm SSS} &= \frac{G\lambda_{1}m_{2}}{3m_{1}^{3}c^{7}}\left[ S_{1}^{ij}S_{1}^{km}S_{1}^{lm} v_{12}^{n} \partial^{1}_{jkln}\left( \frac{1}{r_{12}} \right) - v_{12}^{n}S_{1}^{nj}S_{1}^{km}S_{1}^{lm} \partial^{1}_{ijkl}\left( \frac{1}{r_{12}} \right) \right] \nn\\
	&\quad + \frac{G\lambda_{2}}{3m_{2}^{2}c^{7}}\left[ S_{2}^{ij}S_{2}^{km}S_{2}^{lm} v_{12}^{n} \partial^{1}_{jkln}\left( \frac{1}{r_{12}} \right) - v_{12}^{n}S_{2}^{nj}S_{2}^{km}S_{2}^{lm} \partial^{1}_{ijkl}\left( \frac{1}{r_{12}} \right) \right] \nn\\
	&\quad -\frac{G\kappa_{1}m_{2}}{2m_{1}^{3}c^{7}} S_{1}^{ij}S_{1}^{km}S_{1}^{lm} v_{12}^{n} \partial^{1}_{jkln}\left( \frac{1}{r_{12}} \right) \nn\\
	&\quad + \frac{G\kappa_{1}}{m_{1}^{2}c^{7}}\left[ S_{2}^{ij}S_{1}^{km}S_{1}^{lm} v_{12}^{n} \partial^{1}_{jkln}\left( \frac{1}{r_{12}} \right) - v_{12}^{n}S_{2}^{nj}S_{1}^{km}S_{1}^{lm} \partial^{1}_{ijkl}\left( \frac{1}{r_{12}} \right) \right] \nn\\
	&\quad + \frac{G\kappa_{2}}{m_{1}m_{2}c^{7}}\left[ \frac{1}{2}S_{1}^{ij}S_{2}^{km}S_{2}^{lm} v_{12}^{n} \partial^{1}_{jkln}\left( \frac{1}{r_{12}} \right) - v_{12}^{n}S_{1}^{nj}S_{2}^{km}S_{2}^{lm} \partial^{1}_{ijkl}\left( \frac{1}{r_{12}} \right) \right] \nn\\
	&\quad -\frac{G}{m_{1}^{2}c^{7}} S_{1}^{ij}S_{1}^{km}S_{2}^{lm} v_{12}^{n} \partial^{1}_{jkln}\left( \frac{1}{r_{12}} \right) +\calO(9) \,.
\end{align}
Note the presence of terms devoid of the constants $\kappa$, $\lambda$, coming from the relation $p\leftrightarrow u$. For completeness, we also give the already known leading-order contributions to the equations of motion for lower orders in spin, which will intervene in the wave generation formalism and read
\begin{subequations}
\begin{align}
	\left( a_{1}^{i} \right)_{\rm NS} &= G m_{2} \partial^{1}_{i}\left( \frac{1}{r_{12}} \right) + \calO(2)\,, \\
	\left( a_{1}^{i} \right)_{\rm SO} &= \frac{G m_{2}}{m_{1} c^{3}} \left[ S_{1}^{ij}v_{12}^{n} \partial^{1}_{jn}\left( \frac{1}{r_{12}} \right) - 2 v_{12}^{n} S_{1}^{nj}  \partial^{1}_{ij}\left( \frac{1}{r_{12}} \right) \right] \nn \\
	& \quad + \frac{G}{c^{3}} \left[ 2 S_{2}^{ij}v_{12}^{n} \partial^{1}_{jn}\left( \frac{1}{r_{12}} \right) - 2 v_{12}^{n} S_{2}^{nj}  \partial^{1}_{ij}\left( \frac{1}{r_{12}} \right) \right] + \calO(5) \,, \\
	\left( a_{1}^{i} \right)_{\rm SS} &= \frac{G}{2 m_{1}c^{4}} \left[ \frac{\kappa_{1}m_{2}}{m_{1}}S_{1}^{jl}S_{1}^{kl} + 2 S_{1}^{jl}S_{2}^{kl} + \frac{\kappa_{2}m_{1}}{m_{2}}S_{2}^{jl}S_{2}^{kl}\right] \partial^{1}_{ijk}\left( \frac{1}{r_{12}} \right) + \calO(6) \,.
\end{align}
\end{subequations}
We see that, at leading order, all the dependence on the positions can be expressed with derivatives of $1/r_{12}$. The rule for expanding these derivatives, generating much longer expressions, is given by $\partial^{1}_{L}(1/r_{12}) = (-1)^{\ell}(2\ell-1)!! n_{12}^{<L>}/r_{12}^{\ell+1}$, using the notations introduced in the next section for multi-indices.

As discussed above, since we are working essentially at leading PN order we only need the SS $\calO(3)$ in the precession equation of the spins~\eqref{eq:precessionScalarity}. We obtain
\be
	\left( \frac{\ud S_{1}^{ij}}{\ud t} \right)_{\rm SS} = \frac{2G}{c^{3}} \left[ -\frac{m_{2}\kappa_{1}}{m_{1}} S_{1}^{kl}S_{1}^{l[i}\partial^{1}_{j]k}\left( \frac{1}{r_{12}} \right) - S_{2}^{kl}S_{1}^{l[i}\partial^{1}_{j]k}\left( \frac{1}{r_{12}} \right) + S_{1}^{[i|k}S_{2}^{j]l} \partial^{1}_{kl}\left( \frac{1}{r_{12}} \right) \right] +\calO(5) \,.
\ee
If we translate this for the conserved norm vectors, that obey a precession equation of the form $\ud S_{1}^{i}/\ud t  = \varepsilon_{ijk} \Omega_{1}^{j}S_{1}^{k}$, we find\footnote{Note that the ``SO'' terms here are in fact quadratic-in-spin effects, as shown by the presence of $\kappa_{1}$.}
\be\label{eq:Omega1iS}
	\left( \Omega_{1}^{i} \right)_{\rm SO} = \frac{G \kappa_{1} m_{2}}{m_{1}c^{3}} S_{1}^{k}\partial^{1}_{ik}\left( \frac{1}{r_{12}} \right) + \frac{G}{c^{3}} S_{2}^{k} \partial^{1}_{ik}\left( \frac{1}{r_{12}} \right) + \calO(5) \,.
\ee

Next, we can look for the expression of the conserved orbital energy $E$, such that $\ud E/\ud t =0$ according to the equations of motion and precession. Because of the structure of~\eqref{eq:PNstructA} and~\eqref{eq:PNstructOmega}, we see that the leading order SO and SS terms in the energy contribute in this calculation. They are given by
\begin{subequations}\label{eq:ENSSOSS}
\begin{align}
	\left( E \right)_{\rm NS} &= \frac{m_{1}}{2}v_{1}^{2} + \frac{m_{2}}{2}v_{2}^{2} - \frac{G m_{1}m_{2}}{r_{12}} + \calO(2) \,, \\
	\left( E \right)_{\rm SO} &= -\frac{G m_{2}}{c^{3}} S_{1}^{ij}v_{1}^{j} \partial^{1}_{i}\left( \frac{1}{r_{12}} \right) + \exch + \calO(5) \, \\
	\left( E \right)_{\rm SS} &= -\frac{G}{2 c^{4}} \left[ \frac{\kappa_{1}m_{2}}{m_{1}}S_{1}^{ik}S_{1}^{jk} + S_{1}^{ik}S_{2}^{jk} \right] \partial^{1}_{ij}\left( \frac{1}{r_{12}} \right) + \exch + \calO(6) \,.
\end{align}
\end{subequations}
We find for the cubic-in-spin contributions in the conserved energy
\begin{align}\label{eq:ESSS}
	(E)_{\rm SSS} &= \frac{G\kappa_{1}}{2m_{1}c^{7}} \left[ \frac{m_{2}}{m_{1}}v_{1}^{i}S_{1}^{ij} - v_{2}^{i}S_{2}^{ij} \right] S_{1}^{km}S_{1}^{lm} \partial^{1}_{jkl}\left( \frac{1}{r_{12}} \right) \nn\\
	&\quad + \frac{G}{m_{1}c^{7}} v_{1}^{i}S_{1}^{ij}S_{1}^{km}S_{2}^{lm} \partial^{1}_{jkl}\left( \frac{1}{r_{12}} \right) + \exch + \calO(9) \,.
\end{align}
We notice that the constants $\lambda_{1,2}$ have disappeared from the conserved energy at this order, whereas they are present in the reduced Hamiltonian of~\cite{LS14b} (see appendix~\ref{app:ADMEFT} for the comparison of the results of the two methods).


\section{Leading order cubic-in-spin effects in the gravitational wave energy flux}\label{sec:PNflux}

In this section, we compute the gravitational wave energy flux emitted by an inspiralling compact binary system. We apply the multipolar post-Newtonian wave generation formalism developed in~\cite{BD86,B87} (for a review and additional references, see~\cite{Bliving}). Since we are working in practice at the leading PN order for the cubic-in-spins contributions, we need only the linearized version of this general formalism, which is much simpler as all non-linearities are in fact ignored. Instead of giving the result for the full waveform $h_{ij}^{\rm TT}$ (or, equivalently, the two polarizations $h_{+}$ and $h_{\times}$, or the modes $h_{\ell m}$ of a spin-weighted spherical harmonics decomposition), which are very long expressions, we focus here only on the energy loss rate that governs, together with the orbital energy, the phasing of circular orbits. We also focus on circular, spin-aligned orbits and not quasi-circular orbits as explained in section~\ref{subsec:spinaligned}. We leave the presentation of the results for the full waveform and the reduction to quasi-circular precessing orbits for future work.

We introduce some additional notations: we use capital letters to indicate multi-indices, e.g. $a_{L}=a_{i_{1}}\dots a_{i_{\ell}}$ for a vector $a_{i}$, and we use the brackets $<>$ or a hat $\hat{a}_{L}$ to indicate the symmetric and trace-free (STF) projection. Dots $\dot{a}$ will represent time derivatives, and an exponent in parenthesis $a^{(n)}$ will mean the $n$th time derivative. A compendium of useful formulas for STF tensors can be found in appendix A of~\cite{BD86}.


\subsection{Energy flux and source multipole moments at leading order}\label{subsec:sourcemoments}

The general expression of the total energy flux emitted is expressed in terms of STF radiative multipolar moments $U_{L}$, $V_{L}$ according to~\cite{Thorne80}
\begin{equation}\label{eq:generalflux}
\mathcal{F} = \sum_{\ell = 2}^{+ \infty} \frac{G}{c^{2\ell+1}}\,\left[ \frac{(\ell+1)(\ell+2)}{(\ell-1) \ell \, \ell! (2\ell+1)!!} U_L^{(1)} U_L^{(1)} + \frac{4\ell (\ell+2)}{c^2(\ell-1) (\ell+1)!  (2\ell+1)!!} V_L^{(1)}V_L^{(1)}\right]\,.
\end{equation}
By applying the multipolar post-Minkowskian algorithm introduced in~\cite{BD86,B87}, these radiative moments $U_{L}$, $V_{L}$ are related to a set of source and gauge moments $I_{L}$, $J_{L}$, $W_{L}$, $X_{L}$, $Y_{L}$, $Z_{L}$ by formulas that include non-linear terms, instantaneous as well as hereditary. Since we are working at higher order in spin but at leading PN order, we will need only the leading order of these formulas, given simply by  $U_{L} = I_{L}^{(\ell)}$, $V_{L} = J_{L}^{(\ell)}$, and in the sum only the quadrupolar term $\ell=2$ will contribute as
\be\label{eq:fluxIJ}
	\left( \calF \right)_{\rm SSS} = \frac{G}{5 c^{5}}\left[ \dddot{I}_{ij}\dddot{I}_{ij} + \frac{16}{9 c^{2}}\dddot{J}_{ij}\dddot{J}_{ij} \right]_{\rm SSS} + \calO(9) \,.
\ee

The expressions of the source moments as integrals over space are obtained by matching of the post-Newtonian, near-zone expansion of the field to its multipolar, far-zone expansion~\cite{B98mult}. They read in general
\begin{subequations}\label{eq:defILJL}
\begin{align}
	I_L(t) &= \mathop{\mathrm{FP}}_{B=0}\,\int \ud^3\mathbf{x}\,\left(r/r_0\right)^B \int^1_{-1} \ud z\left\{\delta_\ell\,\hat{x}_L\,\Sigma -\frac{4(2\ell+1)}{c^2(\ell+1)(2\ell+3)} \,\delta_{\ell+1} \,\hat{x}_{iL} \,\Sigma_i^{(1)}\right.\nn\\
	&\qquad\quad \left. +\frac{2(2\ell+1)}{c^4(\ell+1)(\ell+2)(2\ell+5)} \,\delta_{\ell+2}\,\hat{x}_{ijL}\Sigma_{ij}^{(2)}\right\}(\mathbf{x},t+z\,r/c)\,,\\
	J_L(t) &= \mathop{\mathrm{FP}}_{B=0}\,\varepsilon_{ab<i_\ell} \int \ud^3 \mathbf{x}\,\left(r/r_0\right)^B \int^1_{-1} \ud z\left\{\vph{\frac{1}{1}} \delta_\ell\,\hat{x}_{L-1>a} \,\Sigma_b \right. \nn\\
	&\qquad\quad \left. -\frac{2\ell+1}{c^2(\ell+2)(2\ell+3)} \,\delta_{\ell+1}\,\hat{x}_{L-1>ac} \,\Sigma_{bc}^{(1)}\right\} (\mathbf{x},t+z\,r/c)\,,
\end{align}\end{subequations}
where $\Sigma = (\tau^{00}+\tau^{ii})/c^{2}$, $\Sigma_{i} = \tau^{0i}/c$, $\Sigma_{ij} = \tau^{ij}$ and $\tau^{\mu\nu}$ is the PN expansion of the stress-energy pseudo-tensor, source of the Einstein equations in harmonic coordinates~\eqref{eq:Einsteinharmonic}. $\mathrm{FP}_{B=0}$ stands for a regularization by analytic continuation in the complex plane for $B$, keeping only the finite part in $B=0$. The scale $r_{0}$ is an associated regularization constant. The integration on the intermediate variable $z$, with the weighting function $\delta_{\ell}(z) = a_{\ell}(1-z^{2})^{\ell}$ and $a_{\ell} = (2\ell+1)!!/2^{\ell+1}\ell!$ a normalization constant, can be written as a post-Newtonian expansion according to
\be\label{eq:intdeltal}
	\int^1_{-1} dz~ \delta_\ell(z) \,\Sigma(\mathbf{x},t+z\,r/c) = \sum_{k=0}^{+\infty}\,\frac{(2\ell+1)!!}{(2k)!!(2\ell+2k+1)!!}\,\left(\frac{r}{c}\right)^{2k}\!\Sigma^{(2k)}(\mathbf{x},t)\,.
\ee
At leading order, we can keep only the first term in this expansion, and identify the quantities $\Sigma$, $\Sigma_{i}$ with the matter sources $\sigma$, $\sigma_{i}$ given in~\eqref{eq:sigmaLO} and~\eqref{eq:sigmaiLO} (and the $\Sigma_{ij}$ part will not contribute). The above expressions simplify to
\begin{subequations}\label{eq:leadingILJL}
\begin{align}
	I_L &= \int \ud^3\mathbf{x}\, \left[ \hat{x}_L\,\sigma -\frac{4(2\ell+1)}{c^2(\ell+1)(2\ell+3)} \,\hat{x}_{iL} \,\sigma_i^{(1)}\right] \,, \label{eq:leadingIL} \\
	J_L &= \varepsilon_{ab<i_\ell} \int \ud^3 \mathbf{x} \, \hat{x}_{L-1>a} \,\sigma_b \label{eq:leadingJL} \,,
\end{align}\end{subequations}
where we removed the FP regularization as the integrands now have a compact support. For odd orders in spin, the second term of eq.~\eqref{eq:leadingIL} is of the same order as the first term, and the mass moments $I_{L}$ are 1PN order smaller than the current moments $J_{L}$. Mass and current moments enter at the same level in the expansion of the flux~\eqref{eq:generalflux} in this case.

From the expressions of the leading order $\sigma$, $\sigma_{i}$ given in~\eqref{eq:sigmaLO} and~\eqref{eq:sigmaiLO}, it is straightforward to derive the leading order expression of $I_{L}$, $J_{L}$ for any multipolar order. Although we will need only the quadrupole moments for our purpose, and although we will see that there is no leading order cubic-in-spin contribution to the quadrupoles, we give here the general result for reference. We obtain for the non-spinning, spin-orbit and spin-spin part of the moments
\begin{subequations}\label{eq:ILJLNSSOSS}
\begin{align}
	\left(I_L\right)_{\rm NS} &= m_{1}y_{1}^{<L>} + \exch + \calO(2) \,, \\
	\left(J_L\right)_{\rm NS} &= y_{1}^{a}v_{1}^{b}\varepsilon^{ab<i_\ell} y_{1}^{L-1>} + \exch + \calO(2) \,, \\
	\left(I_L\right)_{\rm SO} &= \frac{2\ell}{c^{3}(\ell+1)} \left[ \ell v_{1}^{a}S_{1}^{b}\varepsilon^{ab<i_\ell} y_{1}^{L-1>} - (\ell-1)y_{1}^{a}S_{1}^{b}\varepsilon^{ab<i_\ell} v_{1}^{i_{\ell-1}}y_{1}^{L-2>} \right] + \exch + \calO(5) \,, \\
	\left(J_L\right)_{\rm SO} &= \frac{\ell+1}{2 c} S_{1}^{<i_{\ell}}y_{1}^{L-1>} + \exch + \calO(3) \,, \\
	\left(I_L\right)_{\rm SS} &= -\frac{\ell(\ell-1)\kappa_{1}}{2 m_{1}c^{4}}  S_{1}^{<i_{\ell}}S_{1}^{i_{\ell-1}}y_{1}^{L-2>} + \exch + \calO(6) \,, \\
	\left(J_L\right)_{\rm SS} &= \frac{(\ell-1)\kappa_{1}}{2 m_{1}c^{4}} \left[ 2 v_{1}^{a}S_{1}^{b}\varepsilon^{ab<i_\ell} S_{1}^{i_{\ell-1}}y_{1}^{L-2>} - (\ell-2)y_{1}^{a}v_{1}^{b}\varepsilon^{ab<i_\ell} S_{1}^{i_{\ell-1}}S_{1}^{i_{\ell-2}}y_{1}^{L-3>} \right] \nn\\ &\quad+ \exch + \calO(6) \,.
\end{align}\end{subequations}
For the cubic-in-spin moments, the presence of three derivatives of the Dirac delta function in $\sigma^{\rm SSS}$, $\sigma^{\rm SSS}_{i}$ (and four derivatives for $\dot{\sigma}^{\rm SSS}_{i}$) leads to the vanishing of the leading order contribution to both the mass and current quadrupoles. For $\ell \geq 3$, we get\footnote{The second term in~\eqref{eq:ILSSS} is present only for $\ell\geq4$}
\begin{subequations}\label{eq:ILJLSSS}
\begin{align}
	\left(I_L\right)_{\rm SSS} &= \frac{\ell(\ell-1)(\ell-2)\lambda_{1}}{3(\ell+1) m_{1}^{2}c^{7}} \left[ -\ell v_{1}^{a}S_{1}^{b}\varepsilon^{ab<i_\ell} S_{1}^{i_{\ell-1}} S_{1}^{i_{\ell-2}} y_{1}^{L-3>} \right. \nn\\ 
	&\qquad \qquad \left.+ (\ell-3)y_{1}^{a}S_{1}^{b}\varepsilon^{ab<i_\ell} v_{1}^{i_{\ell-1}} S_{1}^{i_{\ell-1}} S_{1}^{i_{\ell-2}} v_{1}^{i_{\ell-3}}y_{1}^{L-4>}  \right] + \exch + \calO(6) \,, \label{eq:ILSSS}\\
	\left(J_L\right)_{\rm SSS} &= -\frac{(\ell+1)(\ell-1)(\ell-2)\lambda_{1}}{12 m_{1}^{2}c^{5}} S_{1}^{<i_{\ell}} S_{1}^{i_{\ell-1}} S_{1}^{i_{\ell-2}} y_{1}^{L-3>} + \exch + \calO(6) \,.
\end{align}\end{subequations}
and we have $(I_{ij})_{\rm SSS} = \calO(9)$, $(J_{ij})_{\rm SSS} = \calO(7)$. Thus, there is no cubic-in-spin direct contribution in the moments entering eq.~\eqref{eq:fluxIJ} at the order we are considering. Nevertheless, cubic-in-spin contributions do enter eq.~\eqref{eq:fluxIJ} indirectly through the following channels: (i) the SSS $\calO(7)$ terms in the acceleration~\eqref{eq:a1iScube} when replacing the time derivatives of the Newtonian NS quadrupole $\dddot{I}_{ij}^{\rm NS} = 6m_{1}v_{1}^{<i}a_{1}^{j>} + 2m_{1}y_{1}^{<i}\dot{a}_{1}^{j>} + \exch $; (ii) the SO $\calO(3)$ and the SS $\calO(4)$ terms in the acceleration and spin precession in the time derivatives of the SO and SS moments; (iii) the SO $\calO(3)$ terms in the center-of-mass conversion of SS $\calO(4)$ terms (see next section); (iv) the SO$\times$SS terms generated when squaring the moments.

By carefully taking into account all these contributions, we obtain a rather long expression for the energy flux $\calF$. We will give it explicitly in the next section after conversion to center-of-mass variables.


\subsection{Results in the center-of-mass frame}

We now present the main results of this paper in the center-of-mass (CM) frame, defined by the vanishing of the integral of motion $G_{i}$, which verifies $\ud G_{i}/\ud t = P_{i}$ with $P_{i}$ the conserved linear momentum.

We introduce standard notations for variables in the CM frame, as in previous works~\cite{FBB06,BMFB13}. We define $\nu = m_{1}m_{2}/m^{2}$, $\delta = (m_{1}-m_{2})/m$, $ \bm{x} = r \bm{n} = \bfy_{1} - \bfy_{2}$ the separation of the two bodies, and $\bm{v} = \bm{v}_{1} - \bm{v}_{2}$ their relative velocity. From the conserved norm spin vectors $\bfS_{1}$, $\bfS_{2}$ we define $\bfS \equiv \bfS_{1} + \bfS_{2}$ and $\bfSigma \equiv m(\bfS_{2}/m_{2}-\bfS_{1}/m_{1})$. We also find that the final expressions are more compact if we use $\kappa_{+} = \kappa_{1}+\kappa_{2}$, $\kappa_{-} = \kappa_{1}-\kappa_{2}$ and similarly for $\lambda_{+}$, $\lambda_{-}$. For any vectors $\bf{a,b,c}$ we denote the mixed product by $(a,b,c) = \varepsilon_{ijk}a^{i}b^{j}c^{k}$, and the scalar product by $(ab) = a^{i}b^{i}$.

By writing
\be\label{eq:defzi}
	(y_{1}^{i})_{\rm CM} = \frac{m_{2}}{m} x^{i} + z^{i} \,, \quad (y_{2}^{i})_{\rm CM} = -\frac{m_{1}}{m} x^{i} + z^{i}
\ee
and similarly for the velocities, with $z^{i} = \calO(2)$ and $m = m_{1}+m_{2}$ the total mass, 
imposing the vanishing of $G_{i}$ fixes the vector $z^{i}$. There is no leading order SS $\calO(4)$ term, but there is a SO $\calO(3)$ contribution that we have to take into account, given by
\be\label{eq:ziSO}
	(z^{i})_{\rm SO} = \frac{\nu}{m c^{3}} \varepsilon_{ijk} v^{j}\Sigma^{k} + \calO(5) \,.
\ee
It turns out that there is no SSS $\calO(7)$ additional contribution to $G_{i}$ and $z^{i}$, i.e. we get readily $\ud^{2} G_{i}/\ud t^{2} = 0$ from the SO $\calO(3)$ expression of $G_{i}$ and the equations of motion and precession. In fact, when converting Newtonian quantities to the CM frame, one can ignore the highest-order contributions in~\eqref{eq:defzi} anyway (as explained for instance in~\cite{BMFB13}). The transformation rule~\eqref{eq:defzi}-\eqref{eq:ziSO} will therefore be sufficient for our purposes. 

The CM conversion of the orbital energy given in~\eqref{eq:ENSSOSS}-\eqref{eq:ESSS} yields
\begin{align}
	(E)_{SSS} &= \frac{G\nu}{4 m c^{7} r^{4}} \left\{ (n,v,S) \left[(nS)^2 \left(-30 \kappa_{+}-60\right) + (nS) (n\Sigma) \left(-30 \delta  \kappa_{+}-60 \delta +30 \kappa_{-}\right) \right.\right. \nn\\
	& \qquad\qquad\quad \left. + (n\Sigma)^2 \left(15 \delta  \kappa_{-}-15 \kappa_{+} + \nu \left(30 \kappa_{+}+60\right)\right) + S^{2} \left(6 \kappa_{+}+12\right) \right. \nn\\
	& \qquad\qquad\quad \left. + (S\Sigma) \left(6 \delta  \kappa_{+}+12 \delta -6 \kappa_{-}\right) + \Sigma^{2} \left(3 \kappa_{+}-3 \delta  \kappa_{-} + \nu \left(-6 \kappa_{+}-12\right)\right)\right] \nn\\
	& \qquad\qquad\quad + (n,v,\Sigma) \left[ (nS) (n\Sigma) \left(30 \delta  \kappa_{-}-30 \kappa_{+}-60 + \nu \left(120 \kappa_{+}+240\right)\right) \right. \nn\\
	& \qquad\qquad\quad\qquad\qquad\quad \left. + (n\Sigma)^2 \left(15 \kappa_{-}-15 \delta  \kappa_{+} + \nu \left(30 \delta  \kappa_{+}+60 \delta -60 \kappa_{-}\right)\right) \right. \nn\\
	& \qquad\qquad\quad\qquad\qquad\quad \left. + (nS)^2 \left(-30 \delta  \kappa_{+}-60 \delta\right) + S^{2} \left(6 \delta  \kappa_{+}+12 \delta\right) \right. \nn\\
	& \qquad\qquad\quad\qquad\qquad\quad \left. + (S\Sigma) \left(-6 \delta  \kappa_{-}+6 \kappa_{+}+12 + \nu \left(-24 \kappa_{+}-48\right)\right) \right. \nn\\
	& \qquad\qquad\quad\qquad\qquad\quad \left. + \Sigma^{2} \left(3 \delta  \kappa_{+}-3 \kappa_{-} + \nu \left(-6 \delta  \kappa_{+}-12 \delta +12 \kappa_{-}\right)\right)\right] \nn\\
	& \qquad\qquad\quad \left.+ (v,S,\Sigma) \left[(nS) \left(-6 \delta  \kappa_{+}-12 \delta -6 \kappa_{-}\right) + (n\Sigma) \left(-12 + \nu \left(12 \kappa_{+}+24\right)\right) \right] \right\} \,.
\end{align}
In the general case, for general orbits and when keeping the constants $\kappa_{\pm}$, $\lambda_{\pm}$, the emitted energy flux is a long expression. We split it according to
\begin{align}
	(\calF)_{\rm SSS} &= \frac{G^{3}m \nu^{2}}{15 c^{12}r^{7}} \left[ \vph{\frac{Gm}{r}} (n,v,S)f^{3}_{nvS} + (n,v,\Sigma)f^{3}_{nv\Sigma} + (n,S,\Sigma)f^{3}_{nS\Sigma} + (v,S,\Sigma)f^{3}_{vS\Sigma} \right. \nn\\ & \qquad\qquad\quad + \left. \frac{G m}{r} \left( (n,v,S)f^{4}_{nvS} + (n,v,\Sigma)f^{4}_{nv\Sigma} + (n,S,\Sigma)f^{4}_{nS\Sigma} + (v,S,\Sigma)f^{4}_{vS\Sigma}  \right) \right] \,,
\end{align}
where the coefficients are given by
\begin{align}
	f^{3}_{nvS} &= (nS)^2 (nv)^2 \left(-1440 \kappa_{-}^2-1440 \kappa_{+}^2-10980 \kappa_{+}+120 \lambda_{+}-16920\right) \nn\\
	& \quad + (nS) (n\Sigma) (nv)^2 \left(-1440 \delta  \kappa_{-}^2-1440 \delta  \kappa_{+}^2-13290 \delta  \kappa_{+}+120 \delta  \lambda_{+}-21540 \delta \right.  \nn\\
	& \quad\qquad\qquad\qquad\qquad \left. +2880 \kappa_{-} \kappa_{+}+36630 \kappa_{-}-120 \lambda_{-}\right) \nn\\
	& \quad + (n\Sigma)^2 (nv)^2 \left(1440 \delta  \kappa_{-} \kappa_{+}+19470 \delta  \kappa_{-}-60 \delta  \lambda_{-}-720 \kappa_{-}^2-720 \kappa_{+}^2-19470 \kappa_{+} \right.  \nn\\
	& \quad\qquad\qquad\qquad \left.+60 \lambda_{+}-5520 + \nu \left(1440 \kappa_{-}^2+1440 \kappa_{+}^2+15600 \kappa_{+}-120 \lambda_{+}+26160\right)\right) \nn\\
	& \quad + (nv)^2 S^{2} \left(2256 \kappa_{+}+168 \lambda_{+}+3504\right) \nn\\
	& \quad + (nv)^2 (S\Sigma) \left(2608 \delta  \kappa_{+}+168 \delta  \lambda_{+}+4208 \delta -14300 \kappa_{-}-168 \lambda_{-}\right) \nn\\
	& \quad + (nS) (nv) (vS) \left(-216 \kappa_{-}^2-216 \kappa_{+}^2+3504 \kappa_{+}-960 \lambda_{+}+13632\right) \nn\\
	& \quad + (n\Sigma) (nv) (vS) \left(-108 \delta  \kappa_{-}^2-108 \delta  \kappa_{+}^2+3372 \delta  \kappa_{+}-480 \delta  \lambda_{+}+10056 \delta +216 \kappa_{-} \kappa_{+} \right.  \nn\\
	& \quad\qquad\qquad\qquad\qquad \left. -12948 \kappa_{-}+480 \lambda_{-} -18\nu \kappa_{-}\right) \nn\\
	& \quad + (vS)^2 \left(144 \kappa_{-}^2+144 \kappa_{+}^2-924 \kappa_{+}-2424\right) \nn\\
	& \quad + (nS)^2 v^{2} \left(360 \kappa_{-}^2+360 \kappa_{+}^2+8484 \kappa_{+}+2160 \lambda_{+}+2568\right) \nn\\
	& \quad + (nS) (n\Sigma) v^{2} \left(360 \delta  \kappa_{-}^2+360 \delta  \kappa_{+}^2+8814 \delta  \kappa_{+}+2160 \delta  \lambda_{+}+3228 \delta -720 \kappa_{-} \kappa_{+} \right.  \nn\\
	& \quad\qquad\qquad\qquad\quad \left. -17034 \kappa_{-}-2160 \lambda_{-}\right) \nn\\
	& \quad + (n\Sigma)^2 v^{2} \left(-360 \delta  \kappa_{-} \kappa_{+}-8682 \delta  \kappa_{-}-1080 \delta  \lambda_{-}+180 \kappa_{-}^2+180 \kappa_{+}^2+8682 \kappa_{+} \right.  \nn\\
	& \quad\qquad\qquad\quad \left. +1080 \lambda_{+}+960 + \nu \left(-360 \kappa_{-}^2-360 \kappa_{+}^2-9144 \kappa_{+}-2160 \lambda_{+}-3888\right)\right) \nn\\
	& \quad + S^{2} v^{2} \left(-2400 \kappa_{+}-432 \lambda_{+}-2208\right) \nn\\
	& \quad + (S\Sigma) v^{2} \left(-2460 \delta  \kappa_{+}-432 \delta  \lambda_{+}-2328 \delta +6696 \kappa_{-}+432 \lambda_{-}\right) \nn\\
	& \quad + (nS) (nv) (v\Sigma) \left(-108 \delta  \kappa_{-}^2-108 \delta  \kappa_{+}^2+1440 \delta  \kappa_{+}-480 \delta  \lambda_{+}+6192 \delta +216 \kappa_{-} \kappa_{+} \right.  \nn\\
	& \quad\qquad\qquad\qquad\qquad \left. -1392 \kappa_{-}+480 \lambda_{-} + \nu \left(18 \kappa_{-}\right)\right) \nn\\
	& \quad + (n\Sigma) (nv) (v\Sigma) \left(216 \delta  \kappa_{-} \kappa_{+}-7824 \delta  \kappa_{-}+480 \delta  \lambda_{-}-108 \kappa_{-}^2-108 \kappa_{+}^2+7824 \kappa_{+} \right.  \nn\\
	& \quad\qquad\qquad\qquad\quad \left. -480 \lambda_{+}+3296 + \nu \left(216 \kappa_{-}^2+216 \kappa_{+}^2-6120 \kappa_{+}+960 \lambda_{+}-18708\right)\right) \nn\\
	& \quad + (vS) (v\Sigma) \left(144 \delta  \kappa_{-}^2+144 \delta  \kappa_{+}^2-1050 \delta  \kappa_{+}-2676 \delta -288 \kappa_{-} \kappa_{+}+2322 \kappa_{-}\right) \nn\\
	& \quad + (v\Sigma)^2 \left(-144 \delta  \kappa_{-} \kappa_{+}+1224 \delta  \kappa_{-}+72 \kappa_{-}^2+72 \kappa_{+}^2-1224 \kappa_{+}-452 \right.  \nn\\
	& \quad\qquad\qquad \left. + \nu \left(-144 \kappa_{-}^2-144 \kappa_{+}^2+1176 \kappa_{+}+2948\right)\right) \nn\\
	& \quad + (nv)^2 \Sigma^{2} \left(-7326 \delta  \kappa_{-}-84 \delta  \lambda_{-}+7326 \kappa_{+}+84 \lambda_{+}+164 \right.  \nn\\
	& \quad\qquad\qquad\quad \left.  + \nu \left(-2960 \kappa_{+}-168 \lambda_{+}-4948\right)\right) \nn\\
	& \quad + \Sigma^{2}  v^{2} \left(3378 \delta  \kappa_{-}+216 \delta  \lambda_{-}-3378 \kappa_{+}-216 \lambda_{+}-216 + \nu \left(2520 \kappa_{+}+432 \lambda_{+}+2388\right)\right) \nn\\
	f^{3}_{nv\Sigma} &= (nS)^2 (nv)^2 \left(-720 \delta  \kappa_{-}^2-720 \delta  \kappa_{+}^2-5340 \delta  \kappa_{+}+60 \delta  \lambda_{+}-8160 \delta +1440 \kappa_{-} \kappa_{+} \right.  \nn\\
	& \quad\qquad\qquad\qquad \left. -26820 \kappa_{-}-60 \lambda_{-}\right) \nn\\
	& \quad + (nS) (n\Sigma) (nv)^2 \left(2880 \delta  \kappa_{-} \kappa_{+}-7500 \delta  \kappa_{-}-120 \delta  \lambda_{-}-1440 \kappa_{-}^2-1440 \kappa_{+}^2+7500 \kappa_{+} \right.  \nn\\
	& \quad\qquad\qquad\qquad \left. +120 \lambda_{+}-7980 + \nu \left(2880 \kappa_{-}^2+2880 \kappa_{+}^2+25980 \kappa_{+}-240 \lambda_{+}+41880\right)\right) \nn\\
	& \quad + (n\Sigma)^2 (nv)^2 \left(-720 \delta  \kappa_{-}^2-720 \delta  \kappa_{+}^2-3240 \delta  \kappa_{+}+60 \delta  \lambda_{+}-360 \delta +1440 \kappa_{-} \kappa_{+} \right.  \nn\\
	& \quad\qquad\qquad\qquad\quad \left. +3240 \kappa_{-}-60 \lambda_{-} + \nu \left(720 \delta  \kappa_{-}^2+720 \delta  \kappa_{+}^2+7650 \delta  \kappa_{+}-60 \delta  \lambda_{+} \right.\right.  \nn\\
	& \quad\qquad\qquad\qquad\qquad\qquad\qquad\qquad\qquad\quad \left.\left. +12780 \delta -4320 \kappa_{-} \kappa_{+}-14130 \kappa_{-}+180 \lambda_{-}\right)\right) \nn\\
	& \quad + (nv)^2 S^{2} \left(1316 \delta  \kappa_{+}+84 \delta  \lambda_{+}+2128 \delta +12272 \kappa_{-}-84 \lambda_{-}\right) \nn\\
	& \quad + (nv)^2 (S\Sigma) \left(4758 \delta  \kappa_{-}-168 \delta  \lambda_{-}-4758 \kappa_{+}+168 \lambda_{+}+2164 \right.  \nn\\
	& \quad\qquad\qquad\qquad \left. + \nu \left(-5968 \kappa_{+}-336 \lambda_{+}-9884\right)\right) \nn\\
	& \quad + (nS) (nv) (vS) \left(-108 \delta  \kappa_{-}^2-108 \delta  \kappa_{+}^2+1608 \delta  \kappa_{+}-480 \delta  \lambda_{+}+6528 \delta +216 \kappa_{-} \kappa_{+} \right.  \nn\\
	& \quad\qquad\qquad\qquad\qquad \left. +9264 \kappa_{-}+480 \lambda_{-}\right) \nn\\
	& \quad + (n\Sigma) (nv) (vS) \left(216 \delta  \kappa_{-} \kappa_{+}-2580 \delta  \kappa_{-}+480 \delta  \lambda_{-}-108 \kappa_{-}^2-108 \kappa_{+}^2+2580 \kappa_{+} \right.  \nn\\
	& \quad\qquad\qquad\qquad\qquad \left. -480 \lambda_{+}+4824 + \nu \left(-9 \delta  \kappa_{-}+216 \kappa_{-}^2+216 \kappa_{+}^2-6447 \kappa_{+} \right.\right.  \nn\\
	& \quad\qquad\qquad\qquad\qquad\qquad\qquad\qquad\qquad\quad \left.\left. +960 \lambda_{+}-19566\right)\right) \nn\\
	& \quad + (vS)^2 \left(72 \delta  \kappa_{-}^2+72 \delta  \kappa_{+}^2-468 \delta  \kappa_{+}-1224 \delta -144 \kappa_{-} \kappa_{+}-1296 \kappa_{-}\right) \nn\\
	& \quad + (nS)^2 v^{2} \left(180 \delta  \kappa_{-}^2+180 \delta  \kappa_{+}^2+4248 \delta  \kappa_{+}+1080 \delta  \lambda_{+}+1296 \delta -360 \kappa_{-} \kappa_{+} \right.  \nn\\
	& \quad\qquad\qquad\quad \left. +12240 \kappa_{-}-1080 \lambda_{-}\right) \nn\\
	& \quad + (nS) (n\Sigma) v^{2} \left(-720 \delta  \kappa_{-} \kappa_{+}+3552 \delta  \kappa_{-}-2160 \delta  \lambda_{-}+360 \kappa_{-}^2+360 \kappa_{+}^2-3552 \kappa_{+} \right.  \nn\\
	& \quad\qquad\qquad\qquad \left. +2160 \lambda_{+}+996 + \nu \left(-720 \kappa_{-}^2-720 \kappa_{+}^2-17652 \kappa_{+}-4320 \lambda_{+}-6504\right)\right) \nn\\
	& \quad + (n\Sigma)^2 v^{2} \left(180 \delta  \kappa_{-}^2+180 \delta  \kappa_{+}^2+444 \delta  \kappa_{+}+1080 \delta  \lambda_{+}-360 \kappa_{-} \kappa_{+}-444 \kappa_{-}-1080 \lambda_{-} \right.  \nn\\
	& \quad\qquad\qquad\quad \left.  + \nu \left(-180 \delta  \kappa_{-}^2-180 \delta  \kappa_{+}^2-4578 \delta  \kappa_{+}-1080 \delta  \lambda_{+}-1956 \delta +1080 \kappa_{-} \kappa_{+} \right.\right.  \nn\\
	& \quad\qquad\qquad\qquad\quad  \left.\left. +5466 \kappa_{-}+3240 \lambda_{-}\right)\right) \nn\\
	& \quad + S^{2} v^{2} \left(-1236 \delta  \kappa_{+}-216 \delta  \lambda_{+}-1176 \delta -4968 \kappa_{-}+216 \lambda_{-}\right) \nn\\
	& \quad + (S\Sigma) v^{2} \left(-1554 \delta  \kappa_{-}+432 \delta  \lambda_{-}+1554 \kappa_{+}-432 \lambda_{+}-1152 \right.  \nn\\
	& \quad\quad\qquad\qquad \left.  + \nu \left(5064 \kappa_{+}+864 \lambda_{+}+5004\right)\right) \nn\\
	& \quad + (nS) (nv) (v\Sigma) \left(216 \delta  \kappa_{-} \kappa_{+}+4164 \delta  \kappa_{-}+480 \delta  \lambda_{-}-108 \kappa_{-}^2-108 \kappa_{+}^2-4164 \kappa_{+} \right.  \nn\\
	& \quad\qquad\qquad\qquad\qquad \left. -480 \lambda_{+}+1408 + \nu \left(9 \delta  \kappa_{-}+216 \kappa_{-}^2+216 \kappa_{+}^2-2601 \kappa_{+} \right.\right.  \nn\\
	& \quad\qquad\qquad\qquad\qquad\qquad\qquad\qquad\qquad\quad \left.\left. +960 \lambda_{+}-11934\right)\right) \nn\\
	& \quad + (n\Sigma) (nv) (v\Sigma) \left(-108 \delta  \kappa_{-}^2-108 \delta  \kappa_{+}^2+2244 \delta  \kappa_{+}-480 \delta  \lambda_{+}+192 \delta +216 \kappa_{-} \kappa_{+} \right.  \nn\\
	& \quad\qquad\qquad\qquad\qquad \left. -2244 \kappa_{-}+480 \lambda_{-} + \nu \left(108 \delta  \kappa_{-}^2+108 \delta  \kappa_{+}^2-2916 \delta  \kappa_{+}+480 \delta  \lambda_{+} \right.\right.  \nn\\
	& \quad\qquad\qquad\qquad\qquad\qquad\qquad\qquad\qquad\qquad \left.\left. -9144 \delta -648 \kappa_{-} \kappa_{+}+7404 \kappa_{-}-1440 \lambda_{-}\right)\right) \nn\\
	& \quad + (vS) (v\Sigma) \left(-288 \delta  \kappa_{-} \kappa_{+}-66 \delta  \kappa_{-}+144 \kappa_{-}^2+144 \kappa_{+}^2+66 \kappa_{+}-1072 \right.  \nn\\
	& \quad\qquad\qquad\qquad \left.  + \nu \left(-288 \kappa_{-}^2-288 \kappa_{+}^2+2124 \kappa_{+}+5380\right)\right) \nn\\
	& \quad + (v\Sigma)^2 \left(72 \delta  \kappa_{-}^2+72 \delta  \kappa_{+}^2-348 \delta  \kappa_{+}-24 \delta -144 \kappa_{-} \kappa_{+}+348 \kappa_{-} \right.  \nn\\
	& \quad\qquad\qquad \left.  + \nu \left(-72 \delta  \kappa_{-}^2-72 \delta  \kappa_{+}^2+594 \delta  \kappa_{+}+1476 \delta +432 \kappa_{-} \kappa_{+}-1290 \kappa_{-}\right)\right) \nn\\
	& \quad + (nv)^2 \Sigma^{2} \left(720 \delta  \kappa_{+}+84 \delta  \lambda_{+}-252 \delta -720 \kappa_{-}-84 \lambda_{-} \right.  \nn\\
	& \quad\qquad\qquad\quad \left.  + \nu \left(-1668 \delta  \kappa_{+}-84 \delta  \lambda_{+}-2832 \delta +3108 \kappa_{-}+252 \lambda_{-}\right)\right) \nn\\
	& \quad + \Sigma^{2}  v^{2} \left(-312 \delta  \kappa_{+}-216 \delta  \lambda_{+}-36 \delta +312 \kappa_{-}+216 \lambda_{-} \right.  \nn\\
	& \quad\qquad\qquad \left.  + \nu \left(1296 \delta  \kappa_{+}+216 \delta  \lambda_{+}+1296 \delta -1920 \kappa_{-}-648 \lambda_{-}\right)\right) \nn\\
f^{3}_{nS\Sigma} &= (nS) (nv)^3 \left(-3180 \delta  \kappa_{+}-6360 \delta +25380 \kappa_{-}\right) \nn\\
	& \quad + (n\Sigma) (nv)^3 \left(14280 \delta  \kappa_{-}-14280 \kappa_{+}-8220 + \nu \left(6360 \kappa_{+}+12720\right)\right) \nn\\
	& \quad + (nv)^2 (vS) \left(2290 \delta  \kappa_{+}+4580 \delta -20582 \kappa_{-} + \nu \left(-18 \kappa_{-}\right)\right) \nn\\
	& \quad + (nS) (nv) v^{2} \left(-144 \delta  \kappa_{+}-288 \delta -11976 \kappa_{-} + \nu \left(18 \kappa_{-}\right)\right) \nn\\
	& \quad + (n\Sigma) (nv) v^{2} \left(-5916 \delta  \kappa_{-}+5916 \kappa_{+}+1788 + \nu \left(9 \delta  \kappa_{-}+279 \kappa_{+}+606\right)\right) \nn\\
	& \quad + (vS) v^{2} \left(-156 \delta  \kappa_{+}-312 \delta +6492 \kappa_{-}\right) \nn\\
	& \quad + (nv)^2 (v\Sigma) \left(-11436 \delta  \kappa_{-}+11436 \kappa_{+}+4336 + \nu \left(-9 \delta  \kappa_{-}-4571 \kappa_{+}-9070\right)\right) \nn\\
	& \quad + (v\Sigma) v^{2} \left(3324 \delta  \kappa_{-}-3324 \kappa_{+}-224 + \nu \left(312 \kappa_{+}+584\right)\right) \nn\\
f^{3}_{vS\Sigma} &= (nS) (nv)^2 \left(3212 \delta  \kappa_{+}+6424 \delta -13720 \kappa_{-} -18\nu \kappa_{-}\right) \nn\\
	& \quad + (n\Sigma) (nv)^2 \left(-8466 \delta  \kappa_{-}+8466 \kappa_{+}+7480 + \nu \left(-9 \delta  \kappa_{-}-6415 \kappa_{+}-12842\right)\right) \nn\\
	& \quad + (nv) (vS) \left(-1758 \delta  \kappa_{+}-3516 \delta +7650 \kappa_{-} + 18 \nu\kappa_{-}\right) \nn\\
	& \quad + (nS) v^{2} \left(-324 \delta  \kappa_{+}-648 \delta +7008 \kappa_{-}\right) \nn\\
	& \quad + (n\Sigma) v^{2} \left(3666 \delta  \kappa_{-}-3666 \kappa_{+}-2124 + \nu \left(648 \kappa_{+}+1356\right)\right) \nn\\
	& \quad + (nv) (v\Sigma) \left(4704 \delta  \kappa_{-}-4704 \kappa_{+}-3204 + \nu \left(9 \delta  \kappa_{-}+3507 \kappa_{+}+6886\right)\right) \nn\\
f^{4}_{nvS} &= (nS)^2 \left(216 \delta  \kappa_{-} \kappa_{+}+432 \delta  \kappa_{-}-324 \kappa_{-}^2-324 \kappa_{+}^2-96 \kappa_{+}+720 \lambda_{+}-3216\right) \nn\\
	& \quad + (nS) (n\Sigma) \left(-432 \delta  \kappa_{-}^2-432 \delta  \kappa_{+}^2-960 \delta  \kappa_{+}+720 \delta  \lambda_{+}-4512 \delta +864 \kappa_{-} \kappa_{+}+1032 \kappa_{-} \right.  \nn\\
	& \quad\qquad\qquad\qquad \left. -720 \lambda_{-} + \nu \left(-864 \kappa_{-} \kappa_{+}-1728 \kappa_{-}\right)\right) \nn\\
	& \quad + (n\Sigma)^2 \left(432 \delta  \kappa_{-} \kappa_{+}+732 \delta  \kappa_{-}-360 \delta  \lambda_{-}-216 \kappa_{-}^2-216 \kappa_{+}^2-732 \kappa_{+}+360 \lambda_{+}-1872 \right.  \nn\\
	& \quad\qquad\qquad \left. + \nu \left(-216 \delta  \kappa_{-} \kappa_{+}-432 \delta  \kappa_{-}+540 \kappa_{-}^2+540 \kappa_{+}^2+1824 \kappa_{+}-720 \lambda_{+}+5808\right)\right) \nn\\
	& \quad + S^{2} \left(-384 \kappa_{+}-144 \lambda_{+}+96\right) \nn\\
	& \quad + (S\Sigma) \left(-312 \delta  \kappa_{+}-144 \delta  \lambda_{+}+240 \delta -176 \kappa_{-}+144 \lambda_{-}\right) \nn\\
	& \quad + \Sigma^{2} \left(-124 \delta  \kappa_{-}+72 \delta  \lambda_{-}+124 \kappa_{+}-72 \lambda_{+}+1288 + \nu \left(240 \kappa_{+}+144 \lambda_{+}-384\right)\right) \nn\\
f^{4}_{nv\Sigma} &= (nS)^2 \left(-216 \delta  \kappa_{-}^2-216 \delta  \kappa_{+}^2-456 \delta  \kappa_{+}+360 \delta  \lambda_{+}-2208 \delta +432 \kappa_{-} \kappa_{+}+360 \kappa_{-} \right.  \nn\\
	& \quad\qquad\quad \left. -360 \lambda_{-} + \nu \left(-432 \kappa_{-} \kappa_{+}-864 \kappa_{-}\right)\right) \nn\\
	& \quad + (nS) (n\Sigma) \left(864 \delta  \kappa_{-} \kappa_{+}+1284 \delta  \kappa_{-}-720 \delta  \lambda_{-}-432 \kappa_{-}^2-432 \kappa_{+}^2-1284 \kappa_{+}+720 \lambda_{+} \right.  \nn\\
	& \quad\qquad\qquad\qquad \left. -1632 + \nu \left(-432 \delta  \kappa_{-} \kappa_{+}-864 \delta  \kappa_{-}+1080 \kappa_{-}^2+1080 \kappa_{+}^2+3552 \kappa_{+} \right.\right.  \nn\\
	& \quad\qquad\qquad\qquad\qquad\qquad\quad \left.\left. -1440 \lambda_{+}+11424\right)\right) \nn\\
	& \quad + (n\Sigma)^2 \left(-216 \delta  \kappa_{-}^2-216 \delta  \kappa_{+}^2-876 \delta  \kappa_{+}+360 \delta  \lambda_{+}+432 \kappa_{-} \kappa_{+}+876 \kappa_{-}-360 \lambda_{-} \right.  \nn\\
	& \quad\qquad\qquad \left.  + \nu \left(324 \delta  \kappa_{-}^2+324 \delta  \kappa_{+}^2+1320 \delta  \kappa_{+}-360 \delta  \lambda_{+}+3504 \delta -1512 \kappa_{-} \kappa_{+} \right.\right.  \nn\\
	& \quad\qquad\qquad\qquad \left.\left. -3072 \kappa_{-}+1080 \lambda_{-}\right) + \nu^2 \left(432 \kappa_{-} \kappa_{+}+864 \kappa_{-}\right)\right) \nn\\
	& \quad + S^{2} \left(48 \delta  \kappa_{+}-72 \delta  \lambda_{+}+528 \delta +360 \kappa_{-}+72 \lambda_{-}\right) \nn\\
	& \quad + (S\Sigma) \left(-4 \delta  \kappa_{-}+144 \delta  \lambda_{-}+4 \kappa_{+}-144 \lambda_{+}-616 + \nu \left(-336 \kappa_{+}+288 \lambda_{+}-2400\right)\right) \nn\\
	& \quad + \Sigma^{2} \left(160 \delta  \kappa_{+}-72 \delta  \lambda_{+}-160 \kappa_{-}+72 \lambda_{-} \right.  \nn\\
	& \quad\qquad\quad \left. + \nu \left(-120 \delta  \kappa_{+}+72 \delta  \lambda_{+}-672 \delta +440 \kappa_{-}-216 \lambda_{-}\right)\right) \nn\\
f^{4}_{nS\Sigma} &= (nS) (nv) \left(-816 \delta  \kappa_{+}-1632 \delta +336 \kappa_{-}\right) \nn\\
	& \quad + (n\Sigma) (nv) \left(576 \delta  \kappa_{-}-576 \kappa_{+}-1224 + \nu \left(1632 \kappa_{+}+3264\right)\right) \nn\\
	& \quad + (vS) \left(48 \delta  \kappa_{+}+96 \delta -112 \kappa_{-}\right) \nn\\
	& \quad + (v\Sigma) \left(-80 \delta  \kappa_{-}+80 \kappa_{+}+1328 + \nu \left(-96 \kappa_{+}-192\right)\right) \nn\\
f^{4}_{vS\Sigma} &= (nS) \left(648 \delta  \kappa_{+}+1296 \delta -72 \kappa_{-}\right) \nn\\
	& \quad + (n\Sigma) \left(-360 \delta  \kappa_{-}+360 \kappa_{+}-160 + \nu \left(-1296 \kappa_{+}-2592\right)\right) \,.
\end{align}
This energy flux is valid for general orbits at this stage.


\subsection{Spin-aligned circular orbits}\label{subsec:spinaligned}

In this section, we specialize our results to circular orbits in the spin-aligned case and derive the associated phase evolution of the binary. Spin-aligned circular orbits correspond to the case where both spins are aligned with the normal to the orbital plane, and where the orbit has no eccentricity. The situation is then completely analogous to non-spinning systems, the systems evolves slowly, due solely to radiation reaction, and one can deduce the phasing of the binary from the balance equation
\be\label{eq:balance}
	\calF = -\frac{\ud E}{\ud t} \,.
\ee

In the case of misaligned spins, then the orbital plane and the spin vectors will all undergo precession. If one works formally at linear order in the spins, one can still find spherical orbits (precessing, but with a constant separation radius and orbital frequency), and one can treat the components $S_{\ell}$, $\Sigma_{\ell}$, the only ones to enter scalar such as the orbital energy and the emitted energy flux at spin-orbit level, as constants (see the discussion in~\cite{BMB13}). Beyond the linear-in-spin approximation, however, the orbit's separation $r$ and orbital frequency $\omega$ will experience variations on the orbital time scales. It is still possible, in this case, to define so-called quasi-circular orbits, authorizing precession and solving for the variations in $r$ and $\omega$ using averages on one orbit, as was done at leading order for SS terms in~\cite{BFH12}. However, the situation is complicated here by the fact that we would have to take into account the precession effects at SO $\calO(3)$ relative order within the averaging procedure for the SS $\calO(4)$ terms. Another complication arises when trying to build the phasing of the binary from~\eqref{eq:balance}. Precession induces a precessional phase to be added to the signal, and when trying to integrate~\eqref{eq:balance} the components of the spins present in both $E$ and $\calF$ must now be considered varying. We chose to focus here on the spin-aligned phasing and to leave for future work both the study of the more general quasi-circular orbits and of the effect of precession in the phasing of the binary.

We define the unit vectors $\bm{\ell} = \bm{n}\times\bm{\lambda} $, normal to the orbital plane, and $\bm{\lambda}$ such that $\dot{\bm{n}} = \omega\bm{\lambda}$ with $\omega$ being the orbital frequency. We denote by $S_{\ell}$ the components of the spins along $\bm{\ell}$, i.e. $S_{\ell} = \bfS\cdot\bm{\ell}$, which will be constants. For circular orbits, we will have $\bm{v} = r\omega\bm{\lambda}$. We introduce the usual PN parameters $\gamma = G m/r c^{2}$ and $x=(G m \omega/c^{3})^{2/3}$. Consistently with the rest of the paper, we will give only the leading order contributions for each order in spin. We refer the reader to the review~\cite{Bliving} for higher-order corrections.

The equations of motion projected along $\bm{n}$ give $\dot{\bm{v}}\cdot\bm{n} = -r \omega^{2}$, which allows us to relate $r$ to $\omega$, or equivalently $\gamma$ to $x$. We obtain
\be
	\gamma = x \left[ 1 + x g_{\rm NS} + x^{3/2} \frac{g_{\rm SO}}{Gm^{2}} + x^{2} \frac{g_{\rm SS}}{G^{2}m^{4}} + x^{7/2} \frac{g_{\rm SSS}}{G^{3}m^{6}} + \calO(8) \right] \,,
\ee
with the leading order spin corrections
\begin{align}
	g_{\rm SO} &= \frac{5}{3} S_{\ell} + \delta\Sigma_{\ell} + \calO(2) \nn\\
	g_{\rm SS} &= S_{\ell}^2 \left(-\frac{\kappa_{+}}{2}-1\right) + S_{\ell} \Sigma_{\ell} \left(-\frac{\delta  \kappa_{+}}{2}-\delta +\frac{\kappa_{-}}{2}\right) + \Sigma_{\ell}^2 \left(\frac{\delta  \kappa_{-}}{4}-\frac{\kappa_{+}}{4} + \nu \left(\frac{\kappa_{+}}{2}+1\right)\right) + \calO(2) \nn\\
	g_{\rm SSS} &= S_{\ell}^3 \left(-\frac{3 \kappa_{+}}{2}+\lambda_{+}-9\right) + S_{\ell}^2 \Sigma_{\ell} \left(-\frac{5 \delta  \kappa_{+}}{2}+\frac{3 \delta  \lambda_{+}}{2}-14 \delta +3 \kappa_{-}-\frac{3 \lambda_{-}}{2}\right) \nn\\
	& \quad + S_{\ell} \Sigma_{\ell}^2 \left(\frac{13 \delta  \kappa_{-}}{4}-\frac{3 \delta  \lambda_{-}}{2}-\frac{13 \kappa_{+}}{4}+\frac{3 \lambda_{+}}{2}-5 + \nu \left(\frac{11 \kappa_{+}}{2}-3 \lambda_{+}+29\right)\right) \nn\\
	& \quad + \Sigma_{\ell}^3 \left(-\frac{5 \delta  \kappa_{+}}{4}+\frac{\delta  \lambda_{+}}{2}+\frac{5 \kappa_{-}}{4}-\frac{\lambda_{-}}{2} + \nu \left(\delta  \kappa_{+}-\frac{\delta  \lambda_{+}}{2}+5 \delta -\frac{7 \kappa_{-}}{2}+\frac{3 \lambda_{-}}{2}\right)\right) + \calO(2) \,.
\end{align}
The reduction of the orbital energy gives
\be
	E = -\frac{1}{2}m\nu c^{2} x \left[ 1 + x e_{\rm NS} + x^{3/2} \frac{e_{\rm SO}}{Gm^{2}} + x^{2} \frac{e_{\rm SS}}{G^{2}m^{4}} + x^{7/2} \frac{e_{\rm SSS}}{G^{3}m^{6}} + \calO(8) \right] \,,
\ee
where
\begin{align}
	e_{\rm SO} &= \frac{14}{3} S_{\ell} + 2 \delta \Sigma_{\ell} + \calO(2) \nn\\
	e_{\rm SS} &=  S_{\ell}^2 \left(-\kappa_{+}-2\right) + S_{\ell} \Sigma_{\ell} \left(- \delta \kappa_{+} -2 \delta +\kappa_{-}\right) + \Sigma_{\ell}^2 \left(\frac{\delta  \kappa_{-}}{2}-\frac{\kappa_{+}}{2} + \nu \left(\kappa_{+}+2\right)\right) + \calO(2) \nn\\
	e_{\rm SSS} &= S_{\ell}^3 \left(2 \kappa_{+}+4 \lambda_{+}-20\right) + S_{\ell}^2 \Sigma_{\ell} \left(2 \delta  \kappa_{+}+6 \delta  \lambda_{+}-32 \delta +4 \kappa_{-}-6 \lambda_{-}\right) \nn\\ 
	& \quad + S_{\ell} \Sigma_{\ell}^2 \left(5 \delta  \kappa_{-}-6 \delta  \lambda_{-}-5 \kappa_{+}+6 \lambda_{+}-12 + \nu \left(-2 \kappa_{+}-12 \lambda_{+}+68\right)\right) \nn\\
	& \quad + \Sigma_{\ell}^3 \left(-3 \delta  \kappa_{+}+2 \delta  \lambda_{+}+3 \kappa_{-}-2 \lambda_{-} + \nu \left(-2 \delta  \lambda_{+}+12 \delta -6 \kappa_{-}+6 \lambda_{-}\right)\right) + \calO(2) \,,
\end{align}
and one can check that the test-mass limit agrees with the energy per unit mass of a test particle in circular equatorial orbits around a Kerr black hole~\cite{BPT72} (we recall that for black holes, $\kappa_{1,2} = 1$ and $\lambda_{1,2} = 1$.). Incidently, the cubic-in-spin term cancels out at $\calO(7)$ in that limit. The reduction of the emitted flux gives
\be
	\calF = \frac{32 \nu^{2}}{5G} c^{5} x^{5} \left[ 1 + x f_{\rm NS} + x^{3/2} \frac{f_{\rm SO}}{Gm^{2}} + x^{2} \frac{f_{\rm SS}}{G^{2}m^{4}} + x^{7/2} \frac{f_{\rm SSS}}{G^{3}m^{6}} \calO(8) \right] \,,
\ee
with
\begin{align}
	f_{\rm SO} &= -4 S_{\ell} - \frac{5}{4} \delta \Sigma_{\ell} + \calO(2) \nn\\
	f_{\rm SS} &= S_{\ell}^2 \left(2 \kappa_{+}+4\right) + S_{\ell} \Sigma_{\ell} \left(2 \delta  \kappa_{+}+4 \delta -2 \kappa_{-}\right) \nn\\
	& \quad + \Sigma_{\ell}^2 \left(-\delta  \kappa_{-}+\kappa_{+}+\frac{1}{16} + \nu \left(-2 \kappa_{+}-4\right)\right)  + \calO(2) \nn\\
	f_{\rm SSS} &= S_{\ell}^3 \left(-\frac{16 \kappa_{+}}{3}-4 \lambda_{+}+\frac{40}{3}\right) + S_{\ell}^2 \Sigma_{\ell} \left(-\frac{35 \delta  \kappa_{+}}{6}-6 \delta  \lambda_{+}+\frac{73 \delta }{3}-\frac{3 \kappa_{-}}{4}+6 \lambda_{-}\right) \nn\\
	& \quad  + S_{\ell} \Sigma_{\ell}^2 \left(-\frac{35 \delta  \kappa_{-}}{12}+6 \delta  \lambda_{-}+\frac{35 \kappa_{+}}{12}-6 \lambda_{+}+\frac{32}{3} + \nu \left(\frac{22 \kappa_{+}}{3}+12 \lambda_{+}-\frac{172}{3}\right)\right) \nn\\
	& \quad  + \Sigma_{\ell}^3 \left(\frac{67 \delta  \kappa_{+}}{24}-2 \delta  \lambda_{+}-\frac{\delta }{8}-\frac{67 \kappa_{-}}{24}+2 \lambda_{-} + \nu \left(\frac{\delta  \kappa_{+}}{2}+2 \delta  \lambda_{+}-11 \delta +\frac{61 \kappa_{-}}{12}-6 \lambda_{-}\right)\right) \nn\\
	& \quad  + \calO(2) \,.
\end{align}
We checked that in the limit of a test mass around a Kerr black hole, this result agrees with the  one of~\cite{TSTS96}, where the flux is computed in the framework of black hole perturbations.

To illustrate the quantitative importance of the newly computed cubic-in-spin terms, we provide an estimate of their contribution to the phasing of the binary system. There are various ways of deducing the phase evolution from the balance equation~\eqref{eq:balance} and from the PN expressions of $E$ and $\calF$ given above, leading to different approximants (see for instance~\cite{Buonanno+09}). Here we give only the result obtained for the orbital phase as a function of the orbital frequency, $\phi(x)$, by applying the TaylorT2 approach, writing $\ud \phi /\ud x = -\omega (\ud E/\ud x)/\calF$, re-expanding to the required order and integrating analytically term by term. The result is
\be
	\phi(x) = -\frac{x^{-5/2}}{32\nu} \left[ 1 + x \varphi_{\rm NS} + x^{3/2} \frac{\varphi_{\rm SO}}{Gm^{2}} + x^{2} \frac{\varphi_{\rm SS}}{G^{2}m^{4}} + x^{7/2} \frac{\varphi_{\rm SSS}}{G^{3}m^{6}} + \calO(8) \right] \,,
\ee
with
\begin{align}
	\varphi_{\rm SO} &= \frac{235}{6} S_{\ell} + \frac{125}{8}\delta \Sigma_{\ell} + \calO(2) \nn\\
	\varphi_{\rm SS} &= S_{\ell}^2 \left(-25 \kappa_{+}-50\right) + S_{\ell} \Sigma_{\ell} \left(-25 \delta  \kappa_{+}-50 \delta +25 \kappa_{-}\right) \nn\\
	& \quad + \Sigma_{\ell}^2 \left(\frac{25 \delta  \kappa_{-}}{2}-\frac{25 \kappa_{+}}{2}-\frac{5}{16} + \nu \left(25 \kappa_{+}+50\right)\right) + \calO(2) \nn\\
	\varphi_{\rm SSS} &= S_{\ell}^3 \left(\frac{185 \kappa_{+}}{2}-55 \lambda_{+}+515\right) \nn\\
	& \quad + S_{\ell}^2 \Sigma_{\ell} \left(\frac{1105 \delta  \kappa_{+}}{8}-\frac{165 \delta  \lambda_{+}}{2}+\frac{3085 \delta }{4}-\frac{4205 \kappa_{-}}{24}+\frac{165 \lambda_{-}}{2}\right) \nn\\
	& \quad + S_{\ell} \Sigma_{\ell}^2 \left(-\frac{2095 \delta  \kappa_{-}}{12}+\frac{165 \delta  \lambda_{-}}{2}+\frac{2095 \kappa_{+}}{12}-\frac{165 \lambda_{+}}{2}+\frac{24815}{96} \right. \nn\\
	& \qquad\qquad\quad \left. + \nu \left(-275 \kappa_{+}+165 \lambda_{+}-1540\right) \vph{\frac{1}{2}}\right) \nn\\
	& \quad + \Sigma_{\ell}^3 \left(\frac{385 \delta  \kappa_{+}}{6}-\frac{55 \delta  \lambda_{+}}{2}+\frac{55 \delta }{64}-\frac{385 \kappa_{-}}{6}+\frac{55 \lambda_{-}}{2} \right. \nn\\
	& \qquad\qquad \left. + \nu \left(-\frac{365 \delta  \kappa_{+}}{8}+\frac{55 \delta  \lambda_{+}}{2}-\frac{1025 \delta }{4}+\frac{4175 \kappa_{-}}{24}-\frac{165 \lambda_{-}}{2}\right)\right) + \calO(2) \,.
\end{align}
\begin{table*}[h]
\begin{center}
{\small
\begin{tabular}{|r|c|c|c|}
\hline
	LIGO/Virgo &  $10 M_{\odot} + 1.4 M_{\odot}$ & $10 M_{\odot} + 10 M_{\odot}$ \\ \hline \hline
	Newtonian &  $3558.9$ & $598.8$ \\ \hline
	1PN & $212.4$ & $59.1$ \\ \hline
	1.5PN & $-180.9+114.0 \chi_1+11.7 \chi_2$ & $-51.2+16.0 \chi_1+16.0 \chi_2$ \\ \hline
	2PN & $9.8 - 10.5 \chi_{1}^{2} - 2.9 \chi_{1}\chi_{2}$ & $4.0 - 1.1 \chi_{1}^{2} - 2.2 \chi_{1}\chi_{2} - 1.1 \chi_{2}^{2} $ \\  \hline
	2.5PN & \begin{tabular}[c]{@{}c@{}} $-20.0+33.8 \chi_1+2.9 \chi_2$ \\ $ + ( 0.1 \chi_{1} + 0.4 \chi_{1}^{3})$ \end{tabular} & \begin{tabular}[c]{@{}c@{}} $-7.1+5.7 \chi_1+5.7 \chi_2$ \\ $ + (0.05 \chi_{1} + 0.15 \chi_{1}^{3} + 0.05 \chi_{2} + 0.15 \chi_{2}^{3}) $ \end{tabular} \\ \hline
	3PN & $2.3 - 13.2\chi_1 - 1.3\chi_2 + (\mathrm{SS}) $ & $2.2-2.6 \chi_1-2.6 \chi_2 + (\mathrm{SS}) $ \\ \hline
	3.5PN & \begin{tabular}[c]{@{}c@{}} $-1.8+11.1 \chi_1+0.8 \chi_2 + (\mathrm{SS}) $ \\ $ - 0.7 \chi_{1}^{3} - 0.3 \chi_{1}^{2}\chi_{2}$ \end{tabular} &  \begin{tabular}[c]{@{}c@{}} $-0.8+1.7 \chi_1+1.7 \chi_2 + (\mathrm{SS}) $ \\ $- 0.05 \chi_{1}^{3} - 0.2 \chi_{1}^{2}\chi_{2} - 0.2 \chi_{1}\chi_{2}^{2} - 0.05 \chi_{2}^{3}$ \end{tabular}\\ \hline
	4PN & $ (\mathrm{NS}) -8.0 \chi_1 - 0.7 \chi_2 + (\mathrm{SS}) $ & $(\mathrm{NS}) -1.5 \chi_1 - 1.5 \chi_2 + (\mathrm{SS}) $\\ \hline
\end{tabular}
}\end{center}
\caption{Contributions to the number of gravitational-wave cycles $\mathcal{N}_\mathrm{GW} = (\phi_\mathrm{max}-\phi_\mathrm{min})/\pi$, in the spin-aligned case and for circular orbits. For binaries targeted by advanced ground-based detectors LIGO/Virgo, we show the number of cycles accumulated from $\omega_\mathrm{min} = \pi\times 10\,\mathrm{Hz}$ to $\omega_\mathrm{max} = \omega_\mathrm{ISCO}=c^3/(6^{3/2}G m)$. For each compact object we define the dimensionless spin parameter $\chi_A$ by $\mathbf{S}\equiv G m^2 \chi \bm{\ell}$. We give all the other contributions known to date, order by order. The 3.5PN terms cubic in $\chi_{1,2}$ are the new result of this paper. The non-spinning (NS) 4PN terms, and spin-spin (SS) 3.5PN (due to tails) and 4PN terms, are still unknown, but the computation of the 3PN SS terms (not included) has been recently completed~\cite{BFMP15}. Quartic-in-spin contributions would also enter at the 4PN order as well. In the left column, $\chi_{2}$ refers to the spin of the neutron star; since astrophysically realistic values are $\chi_{2}\sim 0.1$ at most, we ignore all contributions $\calO(\chi_{2}^{2})$. At the 2.5PN order, we indicate in parenthesis the leading order contribution of the tidal heating and torquing of the black hole(s).
\label{tab:cycles}}
\end{table*}

In table~\ref{tab:cycles}, we give the contribution of each PN order to the phase evolution of the gravitational wave signal emitted by typical binary systems targeted by the upcoming generation of advanced detectors, from the entry in the band of the detector (taken at $10{\rm Hz}$) to a frequency cutoff taken to be the Schwarzschild ISCO $x=1/6$. Since neutron stars are expected to have dimensionless spin parameters of at most $\chi = S/(G m^{2}) \sim 0.1$, we consider only binary black holes and black hole-neutron star systems, and we ignore all the terms that are at least quadratic in the spin of the neutron star. We also include in the table the leading order phasing contribution due to the tidal heating and torquing of black holes (see e.g.~\cite{TMT97, TP08,CPY12}), which enters at the 2.5PN order with both linear and cubic contributions in spin.

Notice however that the relative importance of successive PN terms in the final phasing does depend on the chosen approximant. More detailed studies~\cite{Nitz+13} found that the 3.5PN next-to-next-to-leading order SO contributions computed in~\cite{MBFB13,BMFB13,BMB13} seem important, as including them improves the agreement between different approximants, but the contributions computed in this paper are somewhat smaller, even if they are of the same formal PN order. Ultimately, the results of this paper could be incorporated in the EOB formalism with spins~\cite{Pan+09,Taracchini+12,Taracchini+13}.


\section*{Acknowledgements}
I am thankful to Luc Blanchet, Alessandra Buonanno, Guillaume Faye and Tanja Hinderer for useful discussions and comments. This work was supported by the NASA grant 11-ATP-046, as well as the NASA grant NNX12AN10G at the University of Maryland College Park. Some of our computations were done using Mathematica\textregistered{} and the symbolic tensor calculus package xAct~\cite{xtensor}.


\appendix
\section{Generalization of the evolution equations to higher multipolar order}\label{app:generalization}
 
In this appendix, we generalize the multipolar formalism of section~\ref{sec:formalism} to an arbitrary multipolar order. This generalization is formal and straightforward, and completely parallels the derivations presented above (see also~\cite{Steinhoff14} for a similar result). It is already present implicitly in the seminal work of Bailey and Israel~\cite{BI75}, but we give here a more explicit presentation. We make contact with the results of Dixon's formalism~\cite{Dixon64,Dixon73,Dixon74,Dixon79} of multipole moments.
 
\subsection{Equations of motion and precession}

The Lagrangian fomalism is not limited to the octupolar coupling we considered, and can be extended to include higher-order derivatives of the Riemann tensor as well. Since an antisymmetrized derivative can be rewritten with a Riemann tensor, we only need to consider symmetrized derivatives~\cite{BI75}. We therefore assume the generalized dependence
\be
	L = L\left[ u^{\mu}, \Omega^{\mu\nu},g_{\mu\nu},R_{\mu\nu\rho\sigma},\dots,\nabla_{(\lambda_{1}\dots\lambda_{n})}R_{\mu\nu\rho\sigma},\dots \right] \,,
\ee
and define the multipole moments
\be
	J^{\lambda_{1}\dots\lambda_{n}\mu\nu\rho\sigma}\equiv \beta_{n} \frac{\partial L}{\partial \nabla_{(\lambda_{1}\dots\lambda_{n})}R_{\mu\nu\rho\sigma}} \,,
\ee
which are symmetric in $\lambda_{1}\dots\lambda_{n}$, have the same symmetries as the Riemann tensor on their last four indices and the additional symmetry $J^{\lambda_{1}\dots\lambda_{n-1}[\lambda_{n}\mu\nu]\rho\sigma} =0$ due to the Bianchi identity. $\beta_{n}$ is a numerical constant.
These $J$ moments have therefore the same symmetries as Dixon's $J$ moments (see below).

We extend~\eqref{eq:resultEOM} to obtain the equation of motion
\be\label{eq:EOMgeneral}
	\frac{\uD p_{\mu}}{\ud\tau} = -\frac{1}{2}R_{\mu\nu\rho\sigma}u^{\nu}S^{\rho\sigma} - \frac{1}{6} J^{\lambda\nu\rho\sigma} \nabla_{\mu} R_{\lambda\nu\rho\sigma} + \sum_{n\geq1} \frac{1}{\beta_{n}} J^{\lambda_{1}\dots\lambda_{n}\tau\nu\rho\sigma}\nabla_{\mu}\nabla_{\lambda_{1}\dots\lambda_{n}}R_{\tau\nu\rho\sigma} \,.
\ee
The scalar condition~\eqref{eq:scalarity} generalizes to
\begin{align}\label{eq:scalaritygeneral}
	0&=p^{\mu}u^{\nu} + S^{\mu\rho}\Omega^{\nu}_{\ph{\nu}\rho} - 2  \frac{\partial L}{\partial g_{\mu\nu}} + \frac{2}{3} R^{\mu}_{\ph{\mu}\lambda\rho\sigma}J^{\nu\lambda\rho\sigma} \nn \\
	& \quad - \sum_{n\geq1} \frac{1}{\beta_{n}} \left[4J^{\lambda_{1}\dots\lambda_{n}\nu\tau\rho\sigma}\nabla_{\lambda_{1}\dots\lambda_{n}}R^{\mu}_{\ph{\mu}\tau\rho\sigma} + n  J^{\lambda_{1}\dots\lambda_{n-1}\nu\tau\xi\rho\sigma}\nabla^{\mu}_{\ph{\mu}\lambda_{1}\dots\lambda_{n-1}}R_{\tau\xi\rho\sigma} \right] \,,
\end{align}
which in turns implies that the precession equation~\eqref{eq:precessionScalarity} becomes (we recall that it is independent of the multipolar coulings in its primary form~\eqref{eq:precessionOmega})
\begin{align}\label{eq:precessiongeneral}
	\frac{\uD S^{\mu\nu}}{\ud \tau} &= 2p^{[\mu}u^{\nu]} + \frac{4}{3} R^{[\mu}_{\ph{\mu}\lambda\rho\sigma}J^{\nu]\lambda\rho\sigma} \nn \\
	& \quad - \sum_{n\geq1} \frac{1}{\beta_{n}} \left[8J^{\lambda_{1}\dots\lambda_{n}[\nu|\tau\rho\sigma|}\nabla_{\lambda_{1}\dots\lambda_{n}}R^{\mu]}_{\ph{\mu}\tau\rho\sigma} + 2 n  J^{\lambda_{1}\dots\lambda_{n-1}[\nu|\tau\xi\rho\sigma}\nabla^{|\mu]}_{\ph{|\mu]}\lambda_{1}\dots\lambda_{n-1}}R_{\tau\xi\rho\sigma} \right] \,.
\end{align}
The stress-energy tensor, however, does not seem to admit an immediate generalization. Additional computations are needed order by order, although they are straightforward.

\subsection{Equivalence with Dixon's equations of evolution at octupolar order}

Dixon developed a covariant and quite general multipolar scheme for extended bodies (summarized in~\cite{Dixon64,Dixon73,Dixon74,Dixon79}), and obtained equations of motion and of precession as a formal serie over multipole moments in the limit of a compact object. He introduces two equivalent sets of multipoles moments, $I^{\lambda_{1}\dots\lambda_{n}\mu\nu}$ and $J^{\lambda_{1}\dots\lambda_{n}\mu\nu\rho\sigma}$, with symmetries
\begin{align}
	I^{\lambda_{1}\dots\lambda_{n}\mu\nu} &= I^{(\lambda_{1}\dots\lambda_{n})(\mu\nu)} \,, \quad  I^{(\lambda_{1}\dots\lambda_{n}\mu)\nu} = 0  \,, \nn\\
	J^{\lambda_{1}\dots\lambda_{n}\mu\nu\rho\sigma} &= J^{(\lambda_{1}\dots\lambda_{n})[\mu\nu][\rho\sigma]} = J^{\lambda_{1}\dots\lambda_{n}\rho\sigma\mu\nu} \,, \nn\\
	J^{\lambda_{1}\dots\lambda_{n}\mu[\nu\rho\sigma]} &= J^{\lambda_{1}\dots\lambda_{n-1}[\lambda_{n}\mu\nu]\rho\sigma} = 0 \,,
\end{align} 
and additional orthogonality relations to a timelike vector used in their definition~\cite{Dixon74}.

The two sets of moments are related by
\begin{align}
	J^{\lambda_{1}\dots\lambda_{n}\mu\nu\rho\sigma} &= \frac{1}{2}\left( I^{\lambda_{1}\dots\lambda_{n}\mu[\rho\sigma]\nu} - I^{\lambda_{1}\dots\lambda_{n}\nu[\rho\sigma]\mu}\right) \,, \nn\\
	I^{\lambda_{1}\dots\lambda_{n}\mu\nu} &= \frac{4(n-1)}{n+1} J^{(\lambda_{1}\dots\lambda_{n-1}|\mu|\lambda_{n})\nu} \,.
\end{align}
Dixon's equations of motion and precession read (eqs.~168-169 in~\cite{Dixon79}, with a minus sign in the first term due to the different convention for the Riemann tensor):
\begin{align}
	\frac{\uD p_{\mu}}{\ud \tau} &= -\frac{1}{2}R_{\mu\nu\rho\sigma}u^{\nu}S^{\rho\sigma} + \frac{1}{2} \sum_{n\geq2}\frac{1}{n!}I^{\lambda_{1}\dots\lambda_{n}\rho\sigma}\nabla_{\mu}g_{\rho\sigma,\lambda_{1}\dots\lambda_{n}}\,, \nn\\
	\frac{\uD S^{\mu\nu}}{\ud \tau} &= 2p^{[\mu}u^{\nu]} + \sum_{n\geq1}\frac{1}{n!} g^{\tau[\mu}I^{|\lambda_{1}\dots\lambda_{n}|\nu]\rho\sigma}g_{\{\tau\rho,\sigma\}\lambda_{1}\dots\lambda_{n}} \,.
\end{align}
Here, the additional notation $\{\tau\rho\sigma\} \equiv \tau\rho\sigma - \rho\sigma\tau + \sigma\tau\rho$ is used, and the $g_{\mu\nu},\lambda_{1}\dots\lambda_{n}$ are ``extensions'' of the metric tensor, as defined by Dixon for instance in the Appendix~2 of~\cite{Dixon74}. The first two extensions (and the only ones we will need) are given by (again, with the opposite sign convention with respect to Dixon's):
\be
	g_{\mu\nu,\rho\sigma} = \frac{2}{3}R_{\mu(\rho\sigma)\nu} \,, \quad g_{\mu\nu,\rho\sigma\tau} = -\nabla_{(\rho}R_{\sigma|\mu|\tau)\nu} \,.
\ee
By using these definitions, expanding all the symmetrizations and making a repeated use of the Bianchi identities, we find agreement with Eqs.~\eqref{eq:EOMgeneral} and~\eqref{eq:precessiongeneral}, identifying Dixon's quadrupole and octupole moments with ours, if we fix the numerical constant for the quadrupolar order at $-6$ and
\be
	\beta_{1} = -12
\ee
at the octupolar order. We adopted these conventional values for these constants throughout the paper.

It would be interesting to investigate this correspondence between the two formalisms at higher multipolar order, and to understand better the relation of this two sets of multipoles which have at the moment different definitions (and which are more general, a priori, in Dixon's formalism). We leave this for future work.


\section{Equivalence with ADM/EFT results for the dynamics}\label{app:ADMEFT}

In this appendix, we compare our results to previous ones obtained for the dynamics in the ADM and EFT approach. A few years ago, the authors of~\cite{HS07,HS08} determined the leading order cubic and quartic in spin reduced Hamiltonians by first studying the case of a test particle in a Kerr background, and then by imposing the completion of the Poincar\'e algebra. Recently, the work~\cite{LS14b} generalized this result (and corrected the quartic in spin sector) to a general compact object with generic $\kappa$, $\lambda$ (which are denoted $C_{\rm{ES}^{2}}$, $C_{\rm{BS}^{3}}$ in~\cite{LS14b} and called Wilson coefficients), by extending the EFT approach for spinning objects~\cite{GR06,Porto06,Levi08,Levi10} to include vertices in the action at higher order in spin. The resulting potential can then be rephrased as a reduced Hamiltonian, like in the ADM formalism.

The comparison of the two results requires to transform the dynamical variables, as the gauge as well as the choice of spin variable are different in the ADM and harmonic formalism. However, the correspondence will be simplified by the fact that we are working at leading order. In line with the notations of~\cite{MBFB13,BMFB13}, we use an overbar for ADM quantities. The contact transformation that relates the harmonic positions to the ADM variables is known to be
\be\label{eq:transfoY}
	\bm{y}_{1} = \bm{Y}_{1}(\ov{\bm{x}},\ov{\bm{p}},\ov{\bm{S}}) = \ov{\bm{x}}_{1} - \frac{1}{2m_{1}^{2}c^{3}} \ov{\bm{p}}_{1}\times\ov{\bm{S}}_{1} + \calO(4) \,,
\ee
and it turns out that there are no SS $\calO(4)$ terms in the transformation\footnote{The NS terms start at $\calO(4)$.}. The transformation of the spin vectors starts at $\calO(4)$ for the SO terms and at $\calO(5)$ for the SS terms. Considering the relative orders required, we see that the simple contact transformation is all we need, and that we can ignore the transformation of spins. For the comparison of the equations of motion, we compute (with $\{\}$ the Poisson bracket)
\begin{subequations}
\begin{align}
	\bm{V}_{1}(\ov{\bm{x}},\ov{\bm{p}},\ov{\bm{S}}) &= \left\{\bm{Y}_{1} , H_{\rm ADM} \right\} \,,\label{eq:transfoV}\\
	\bm{A}_{1}(\ov{\bm{x}},\ov{\bm{p}},\ov{\bm{S}}) &= \left\{\bm{V}_{1} , H_{\rm ADM} \right\} \label{eq:transfoA} \,,
\end{align}
\end{subequations}
and we check that the conversion in ADM variables, using the transformations~\eqref{eq:transfoY} and~\eqref{eq:transfoV}, of the harmonic-coordinates energy and acceleration agree with the ADM Hamiltonian and with~\eqref{eq:transfoA}.


\bibliography{/Users/marsat/Documents/publications/bibliographie/ListeRef}

\end{document}